\documentclass[lettersize,journal]{IEEEtran}
\usepackage{amsmath,amsfonts}
\usepackage{algorithmic}
\usepackage{algorithm}
\usepackage{array}
\usepackage{textcomp}
\usepackage{stfloats}
\usepackage{url}
\usepackage{verbatim}
\usepackage{graphicx}
\usepackage{caption}
\usepackage{subcaption}
\usepackage{cite}
\usepackage{amssymb}
\usepackage{amsthm}
\usepackage[most]{tcolorbox}
\usepackage{booktabs}
\usepackage[table]{xcolor} 
\usepackage{multirow}
\usepackage[utf8]{inputenc}
\usepackage{longtable}
\usepackage{tabularx}

\newtheorem{myDef}{Definition}

\begin{document}
\title{Hyperion: Hierarchical Scheduling for Parallel LLM Acceleration in Multi-tier Networks}

\author{Mulei~Ma,~\IEEEmembership{Graduate~Student~Member,~IEEE,}
        Xinyi~Xu,
        Minrui~Xu,~\IEEEmembership{Member,~IEEE,}
        Zihan~Chen,~\IEEEmembership{Member,~IEEE,}
        Tony~Q.~S.~Quek,~\IEEEmembership{Fellow,~IEEE,}
        and~Yang~Yang,~\IEEEmembership{Fellow,~IEEE}%
\thanks{This research is partially supported by the National Key R\&D Program of China under Grant 2024YFE0200500.}%
\thanks{Mulei Ma is with the IoT Thrust, The Hong Kong University of Science and Technology (Guangzhou), China. (e-mail: mma085@connect.hkust-gz.edu.cn).}%
\thanks{Xinyi Xu is with the National Key Laboratory of Wireless Communications, University of Electronic Science and Technology of China, China. (e-mail: xinyixu@std.uestc.edu.cn).}
\thanks{Minrui Xu is with the School of Computer Science and Engineering, Nanyang Technological University, Singapore 639798, Singapore (e-mail: minrui001@e.ntu.edu.sg).}
\thanks{Zihan~Chen is with the Information System Technology and Design Pillar, Singapore University of Technology and Design, Singapore 487372 (e-mail: zihan\_chen@mymail.sutd.edu.sg).}
\thanks{Tony Q. S. Quek is currently the Full Professor with the Singapore University of Technology and Design (SUTD), Singapore 487372, where he also serves as the Head of the ISTD Pillar, the Sector Lead of the SUTD AI Program, and the Deputy Director of the SUTD-ZJU IDEA. (e-mail: tonyquek@sutd.edu.sg)}
\thanks{Yang Yang is with Shanghai HKU Education Centre, The University of Hong Kong, China.  (email: yang.yang@hku.hk)}%
}

\maketitle
\begin{abstract}
LLMs are increasingly executed at the edge, where limited GPU memory and heterogeneous computation jointly constrain deployment, which motivates model partitioning and request scheduling. In this setting, minimizing latency requires addressing the tight coupling between model placement and request scheduling across heterogeneous nodes, as suboptimal decisions in one domain can negate benefits in the other. In this paper, we propose Hyperion, a hierarchical two-stage framework that coordinates partitioning and scheduling for pipelined LLM inference. Hyperion reduces latency by balancing resources across tiers without requiring model retraining. Leveraging the timescale difference between partitioning and request arrivals, Stage 1 performs offline, inter-tier partitioning via a Hyperion Split with Dynamic Programming (HypSplit-DP) procedure to produce balanced stage times under tier capacity and memory constraints; to adapt to time-varying load, Stage 2 performs online, intra-tier scheduling with a lightweight Hyperion Scheduling for Real-Time (HypSched-RT) that maps each request to a feasible node using real-time estimates of queue length and effective capacity. Experiments with Llama 3 and Phi-3 demonstrate that Hyperion reduces latency by up to 46.9\% (vs. GPipe) and 31.1\% (vs. HEFT). Furthermore, Hyperion exhibits superior scalability for long-sequence generation, maintaining 25.5\% lower latency and higher GPU utilization.

\end{abstract}

\begin{IEEEkeywords}
Large Language Models (LLMs), Distributed Inference, Edge Computing, Hierarchical Scheduling
\end{IEEEkeywords}

\section{Introduction}

\begin{figure*}[htbp]
\centerline{\includegraphics[scale=0.5]{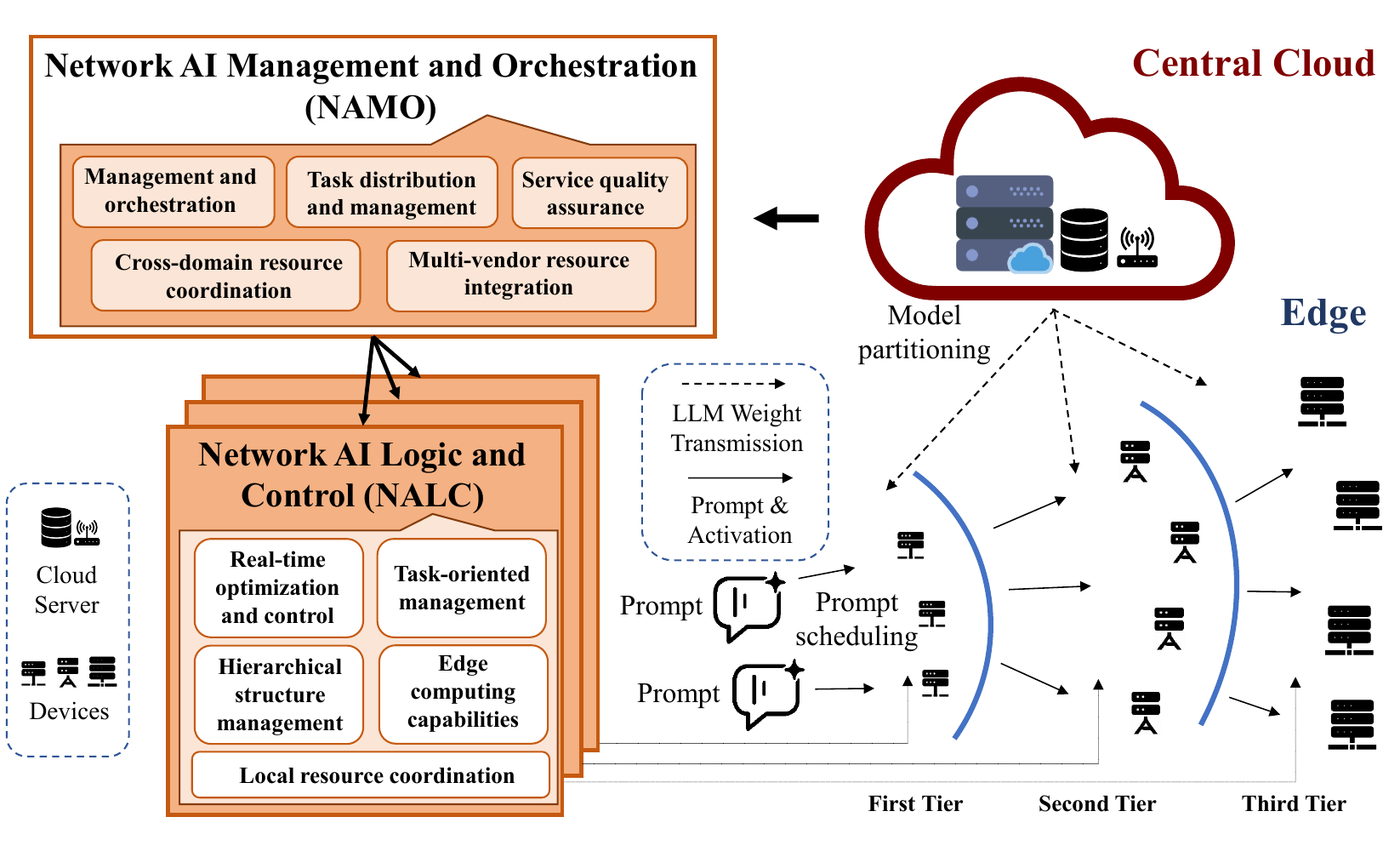}}
\caption{Multi-tier Network Architecture Incorporating Three Edge Tiers for Collaborative LLM Inference and Resource Management}
\label{fig1}
\end{figure*}

Transformer-based large language models (LLMs) have improved performance on a range of natural language processing (NLP) tasks \cite{ref1}. Their adoption is changing computing workloads in cloud and data centers and has prompted research into deployment on edge devices \cite{ref2,chen2024space}. Meanwhile, the development of the Internet of Things (IoT) and 5G has increased demand for on-device AI on smartphones, wearables, and other edge devices. These applications often require low latency and real-time processing. Running LLMs locally can reduce reliance on cloud services and support personalized responses \cite{ref4}.

However, edge devices provide substantially less computing and memory capacity than large LLMs require \cite{ref5,ref6}. For example, NVIDIA reports that the Jetson Nano provides 4 GB of LPDDR4 memory and up to 472 GFLOPS of peak computing performance \cite{refNvidiaNano}. In comparison, GPT-3 contains 175 billion parameters \cite{refGPT3}; storing its model weights alone in FP16 requires approximately 350 GB, calculated as $175\times10^9\times2$ bytes, excluding the KV cache, activations, and other runtime memory. This resource gap motivates collaborative model partitioning across heterogeneous edge and cloud nodes \cite{11314800,ref21}.

To study collaborative LLM inference, we consider a multi-tier network with heterogeneous computing nodes, as shown in Fig. \ref{fig1}. The nodes are arranged in three tiers below a central cloud. Following the terminology in \cite{ref11}, we use Network AI Management and Orchestration (NAMO) for cloud-level coordination and Network AI Logic and Control (NALC) for local control within each tier. NAMO coordinates resources across domains, while NALC manages resources within individual regions. The tiers can follow geographic or organizational boundaries to accommodate local privacy requirements and deployment needs. This architecture supports collaborative LLM inference for latency-sensitive applications.

In this architecture, computing capacity, memory capacity, and network conditions differ across tiers and nodes \cite{refEdgeShard}. Existing approaches address different parts of this deployment problem. On-device inference techniques mainly reduce the execution cost of an individual model instance, while distributed partitioning and computation-offloading methods determine how model execution or inference requests are distributed across devices \cite{ref43}. However, these techniques do not necessarily address the combination of an ordered multi-tier pipeline and multiple candidate execution nodes within each tier.

Hyperion focuses on two related decisions in this setting. The first is inter-tier model partitioning, which assigns contiguous Transformer blocks to heterogeneous tiers under computing and memory constraints. The second is intra-tier request scheduling, which dispatches each arriving prompt or intermediate activation to one of the nodes hosting the corresponding pipeline stage. The partition determines the workload assigned to each tier, whereas the scheduling decision determines the queueing and execution delay experienced by each request. While edge devices are highly sensitive to power consumption, our current approach is strictly latency-centric. Furthermore, pipeline execution can incur idle periods and pipeline overheads. In heterogeneous edge networks, imbalanced stage allocations can increase bottleneck-induced idle time at faster tiers \cite{refDNNPipe}.

Therefore, we propose Hyperion, a two-stage framework for coordinating inter-tier LLM block partitioning and intra-tier request scheduling in a structured multi-tier network. To this end, we first formulate a system-level Min-Max problem aimed at minimizing the inference latency, denoted as $(\mathcal{P}_{0})$, and then introduce Hierarchical Parallel Execution and resource-aware scheduling for LLM Inference on multi-tier Networks (Hyperion), which decomposes the coupled formulation into two less complex, sequential sub-problems. First Stage: Model Partitioning and Inter-tier Allocation Strategy $(\mathcal{P}_{1})$, which is an offline static process that partitions the LLM's blocks across tiers. By considering each tier's computing and memory resources to optimize load balancing, this stage leverages a Hyperion Split with Dynamic Programming (HypSplit-DP) algorithm to determine the partition strategy. Second Stage: Intra-tier Task Scheduling Strategy $(\mathcal{P}_{2})$. Taking the fixed partition from first stage as a given, this stage leverages the Hyperion Scheduling for Real-Time (HypSched-RT) algorithm, which achieves an $O(K_j)$ time complexity via a single linear scan, performing online scheduling of prompts/activations to identify an execution node within each tier. This hierarchical design separates the offline partitioning decision from the online node-selection decision, thereby reducing the computation required at each request arrival. The main contributions of this study are summarized as follows:

\begin{itemize}

\item[1)]
We formulate the inference latency challenge in heterogeneous, memory-constrained multi-tier networks as a Min-Max orchestration problem $(\mathcal{P}_{0})$. We propose Hyperion, a framework that captures cross-tier stage boundaries, memory feasibility, and per-request node selection, and motivates a two-timescale decomposition into the offline subproblem $\mathcal{P}_{1}$ and the online subproblem $\mathcal{P}_{2}$.
\end{itemize}

\begin{itemize}

\item[2)]
For the offline stage, we design a HypSplit-DP algorithm that partitions the LLM’s blocks across tiers by minimizing the maximum profiled stage time subject to memory constraints.
\end{itemize}

\begin{itemize}

\item[3)]
For the online stage, we introduce the lightweight HypSched-RT approach, which dynamically assigns incoming inference requests to the node within each tier by considering real-time device status and availability. The node-selection procedure requires a single scan over the $K_j$ nodes and has $O(K_j)$ per-decision complexity.

\end{itemize}

\begin{itemize}

\item[4)]
We evaluate Hyperion under multiple task loads, model sizes, bandwidth settings, arrival patterns, and output lengths. Results demonstrate that Hyperion reduces end-to-end latency by up to 31.1\% and 46.9\% compared with the HEFT-based and GPipe-based baselines, respectively. At 256 output tokens, it achieves 25.5\% lower latency than the GPipe-based baseline.
\end{itemize}

The remainder of this paper is organized as follows. Section \uppercase\expandafter{\romannumeral2} reviews prior work in distributed LLM inference and resource scheduling. Section \uppercase\expandafter{\romannumeral3} details our system model and the optimization problem formulation. Section \uppercase\expandafter{\romannumeral4} introduces the Hyperion framework, detailing its two-stage scheduling algorithms. Section \uppercase\expandafter{\romannumeral5} presents the experimental evaluation against several baselines. Finally, Section \uppercase\expandafter{\romannumeral6} concludes the paper.

\section{Related work}

\subsection{Distributed Model Partitioning and Pipelined Inference}

Distributed model partitioning enables resource-constrained devices to collaboratively execute models that cannot be accommodated by a single device. DDNN \cite{ref21} partitions a DNN across cloud, edge, and end devices and demonstrates the feasibility of collaborative inference across different network locations. More recently, EdgeShard \cite{refEdgeShard} jointly selects devices and partitions an LLM into shards for collaborative inference over heterogeneous edge devices and cloud servers. EdgeShard focuses on device selection and deployment planning rather than request-level dispatch among replicas of the same pipeline stage. SlimCaching \cite{11395617} considers MoE inference at the edge and caches experts across servers with limited storage. It focuses on expert placement, while Hyperion partitions consecutive Transformer blocks and schedules requests across stage replicas.

Pipeline parallelism provides the execution structure underlying block-partitioned inference. GPipe \cite{ref33} and PipeDream \cite{ref34} establish pipeline-parallel execution and stage partitioning for distributed DNN training. These systems motivate the use of contiguous model stages, but their original formulations address training workloads and do not provide an online scheduler for multi-tier LLM inference. HexGen \cite{ref36} studies generative LLM inference over heterogeneous GPU environments using asymmetric parallelism and topology-aware placement. Its focus is heterogeneous GPU parallelization within a distributed computing environment, rather than hierarchical placement across ordered tiers with replicated execution nodes.

Several recent inference systems are more closely related to the partitioning stage of Hyperion. DNNPipe \cite{refDNNPipe} uses dynamic programming to construct throughput-oriented pipeline plans for DNN inference on asymmetric IoT nodes. Hermes \cite{refHermes} reduces the memory footprint and pipeline stalls of large-model inference through dynamic memory management and parallel model loading. However, DNNPipe optimizes the pipeline plan without considering per-request routing among replicated stage nodes, while Hermes primarily addresses model loading and memory management rather than cross-tier partitioning and queue-aware dispatch.

\subsection{Multi-tier LLM Serving and Resource Scheduling}

Another line of work determines whether and where an inference request or model should be executed in cloud-edge environments. He et al. \cite{ref2} study LLM inference offloading and resource allocation in cloud-edge computing, while Xu et al. \cite{ref19} jointly optimize inference offloading and model caching for small- and large-language-model collaboration. These approaches operate mainly at the model or request-offloading level and do not partition one LLM into an ordered pipeline spanning multiple tiers.

CoEdge \cite{refCoEdge} performs cooperative DNN inference over heterogeneous peer edge devices and adapts its workload partition to computing and network conditions. It addresses heterogeneous distributed inference, but does not study Transformer-block placement across a structured sequence of tiers or request routing among multiple replicas within each tier. RecServe \cite{ref43} deploys progressively capable language models across device, edge, and cloud tiers and recursively offloads requests according to task-specific confidence and dynamically adjusted thresholds. RecServe therefore optimizes model-level cascading and offloading, whereas Hyperion executes different contiguous blocks of the same LLM as pipeline stages.

Hyperion instead assumes a fixed sequence of network tiers in which each tier hosts a contiguous stage of the same LLM and may contain multiple candidate execution nodes. Its decision variables are the stage boundaries across tiers and the node selected for each arriving request within a tier.

\subsection{Single-Node and Decentralized Serving Systems}

LLM serving engines optimize execution at different layers of the serving stack. vLLM \cite{ref44} improves KV-cache memory management through PagedAttention, while Orca \cite{ref45} introduces iteration-level scheduling and selective batching for generative-model serving. PowerInfer \cite{ref46} accelerates single-machine inference through activation-sparsity-aware CPU--GPU execution. These systems optimize memory management, batching, or local execution, but they do not determine how a model is partitioned across network tiers.

Petals \cite{ref47} enables collaborative inference and fine-tuning across Internet-connected peers. Unlike this decentralized setting, Hyperion assumes an administratively managed and ordered multi-tier pipeline with profiled tier capacities.

\subsection{Focused Comparison and Positioning of Hyperion}

\begin{table*}[t]
\centering
\caption{Focused comparison with closely related distributed and multi-tier inference systems.}
\label{tab:focused_comparison}
\scriptsize
\renewcommand{\arraystretch}{1.15}
\begin{tabularx}{\textwidth}{
p{1.45cm}
p{2.65cm}
p{2.65cm}
p{3.05cm}
p{2.75cm}
X}
\toprule
\textbf{System} &
\textbf{Deployment structure} &
\textbf{Model organization} &
\textbf{Deployment-time decision} &
\textbf{Request-time decision} &
\textbf{Primary optimization focus} \\
\midrule

EdgeShard \cite{refEdgeShard} &
Heterogeneous edge devices and cloud servers &
One LLM partitioned into distributed shards &
Joint device selection and model partitioning using dynamic programming &
No per-request selection among multiple replicas of the same stage &
Inference latency or throughput through collaborative model deployment \\

DNNPipe \cite{refDNNPipe} &
Pipeline over asymmetric IoT nodes &
One DNN divided into consecutive pipeline stages &
Dynamic-programming-based pipeline planning &
No request-level routing among replicated stage nodes &
Pipeline throughput and partitioning overhead \\

RecServe \cite{ref43} &
Device--edge--cloud hierarchy &
A complete LLM of different capability at each tier &
Placement of progressively capable LLMs across tiers &
Confidence-based recursive offloading between tiers &
Service quality and communication burden \\

Jupiter \cite{refJupiter} &
Collaborative inference over edge devices &
One generative LLM executed through phase-specific pipelines &
Parallelism planning for prefill and decoding &
Specialized pipeline execution and speculative decoding &
Prefill and autoregressive decoding acceleration \\

Helix \cite{refHelix} &
General heterogeneous GPU and network graph &
Layer placement with flexible per-request pipelines &
MILP- and max-flow-based model placement &
Flow-based request-path selection &
Throughput and latency in heterogeneous GPU clusters \\

Hyperion &
Ordered multi-tier network with multiple candidate nodes per tier &
Contiguous blocks of one LLM assigned as cross-tier stages &
Memory-feasible Min--Max stage partitioning using HypSplit-DP &
Remaining-workload- and memory-aware stage-node selection using HypSched-RT &
End-to-end latency with low online decision overhead \\
\bottomrule
\end{tabularx}
\end{table*}

As summarized in Table \ref{tab:focused_comparison}, existing studies mainly focus on collaborative model partitioning, pipeline planning, model-level offloading, phase-specific inference acceleration, or general heterogeneous-cluster placement. These methods do not simultaneously consider cross-tier stage planning and request-level node selection in an ordered multi-tier pipeline with multiple candidate nodes per tier.

Hyperion combines these two decision levels through a lightweight cross-timescale design. Static pipeline planners fix both the stages and the target node offline, so they cannot react when one node falls behind. Global placement-and-routing formulations do react, but they solve an optimization per request, which is costly on resource-constrained edge controllers. Hyperion sits between the two: the boundaries are solved once, and the per-request decision is a single pass over the $K_j$ nodes of the tier.

\section{System Model and Problem Formulation}

\subsection{Modeling of LLM Structure}

\begin{figure}[t]
\centerline{\includegraphics[scale=0.32]{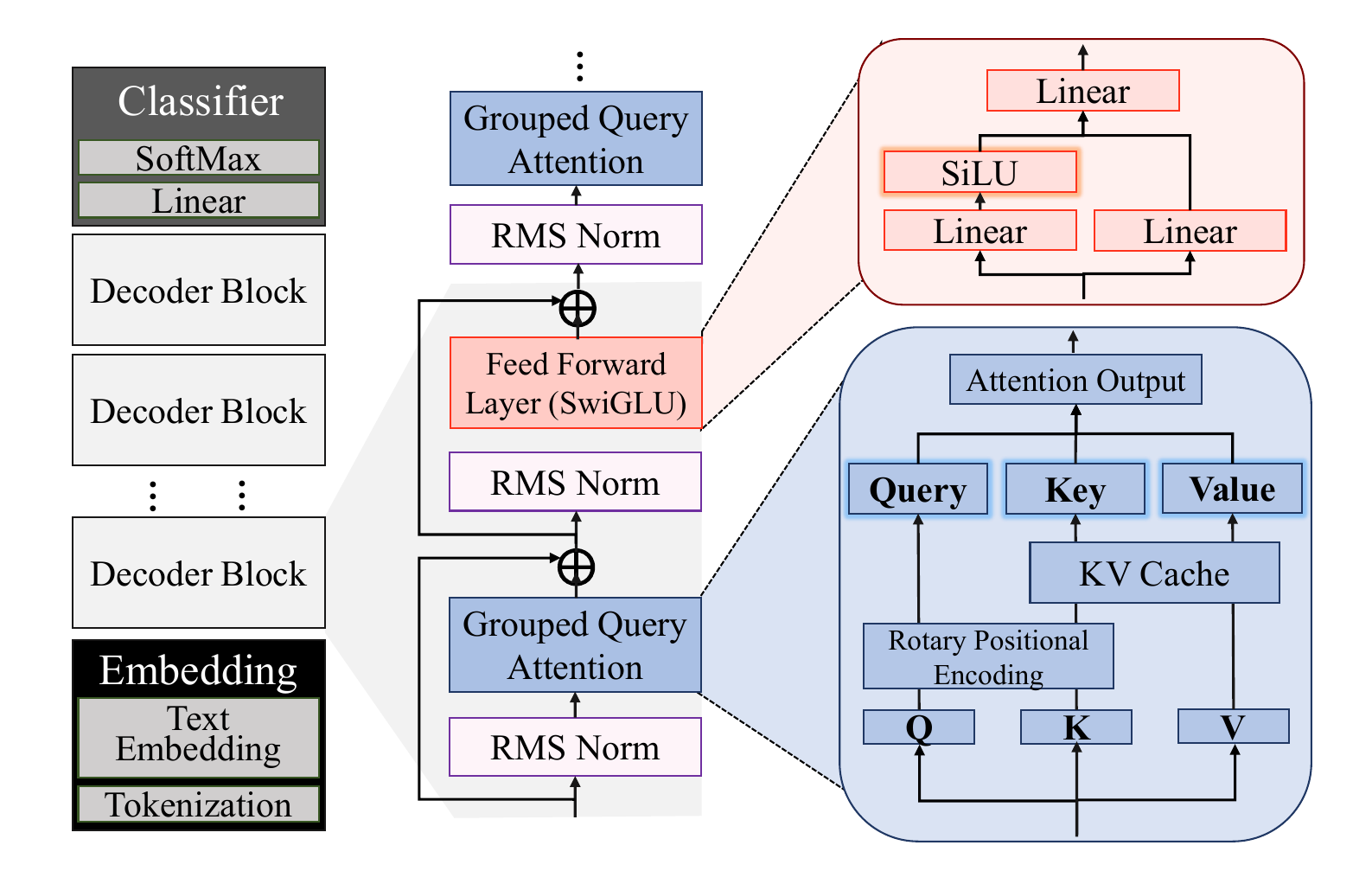}}
\caption{Hierarchical Architecture of a Llama 3, from Decoder Blocks to Detailed GQA and SwiGLU Component Views.}
\label{fig2}
\end{figure}

The inference process of decoder-only LLMs can be represented as a linear Directed Acyclic Graph (DAG). As shown in Fig. \ref{fig2}, we use Llama 3 as an example to illustrate the structure between these decoder blocks. Here, the attention mechanism is implemented as Grouped Query Attention (GQA), which incorporates Rotary Positional Encoding (RoPE). The Feed Forward Network utilizes a SwiGLU gating mechanism, and Root Mean Square Normalization (RMS Norm) is applied before both the attention and feed forward sub layers. These sub layers are sequentially arranged to form a decoder transformer block, and the overall LLM is constructed by stacking a series of these blocks. For partitioning, we treat each decoder block as an indivisible unit.

We abstract an LLM $\mathcal{M}$ as a computation graph composed of $N$ sequential decoder blocks. Let $\mathcal{B}=\left\{B_1, B_2, \ldots, B_N\right\}$ be the ordered set of decoder blocks that constitute the model, where $N \in \mathbb{N}^{+}$ denotes the depth of the model. During the inference process, a strict sequential dependency exists among these blocks, which can be formally expressed as:

\begin{equation}
\forall i \in\{1,2, \ldots, N-1\}, \quad B_i \rightarrow B_{i+1}.
\end{equation}

This implies that the output of block $B_i$ serves as the exclusive input to block $B_{i+1}$. Each decoder block $B_i$ has computing and memory requirements. We define the following two functions to characterize its resource:

\begin{itemize}

  \item \emph{Computing Workload}: We define a function $f: \mathcal{B} \rightarrow \mathbb{R}^{+}$, where $f\left(B_i\right)=f_i$ denotes the total number of Floating Point Operations (FLOPs) required to perform a single forward propagation of module $B_i$.

  \item \emph{Memory Requirements}: We define a function $m: \mathcal{B} \rightarrow \mathbb{R}^{+}$, where $m\left(B_i\right)=m_i$ denotes the memory space, measured in Gigabytes (GB), required on a computing node to load the weight parameters of module $B_i$ and store its intermediate activation values during the forward propagation process.

\end{itemize}

We aggregate the computing workload and memory requirements of the entire model into $N$-dimensional column vectors, respectively:

\begin{equation}
\mathbf{f}=\left[f_1, f_2, \ldots, f_N\right]^T \in \mathbb{R}^{N \times 1} .
\end{equation}

\begin{equation}
\mathbf{m}=\left[m_1, m_2, \ldots, m_N\right]^T \in \mathbb{R}^{N \times 1}.
\end{equation}

\subsection{Multi-tier Heterogeneous Network Architecture}

We consider a multi-tier network composed of $T$ tiers, designed to host the pipelined inference tasks of an LLM. Let $\mathcal{T}=\{1,2, \ldots, T\}$ be the index set of the network tiers. Each tier $j \in \mathcal{T}$ consists of a cluster containing $K_j$ nodes. We represent the set of nodes in tier $j$ as $\mathcal{K}_j=\left\{1,2, \ldots, K_j\right\}$, where $K_j=\left|\mathcal{K}_j\right|$ is the cardinality of nodes in that tier. The set of all nodes across the entire network can thus be expressed as $\mathcal{K}=\bigcup_{j \in \mathcal{T}}\left\{(j, k) \mid k \in \mathcal{K}_j\right\}$.

Each node $(j, k)$ in the network (the $k$-th node within tier $j$) is equipped with independent computing and memory resources. In this work, device heterogeneity is inter-tier. That is, within any given tier $j$, all nodes have the same hardware specification and are treated as homogeneous. The processing speed of a node is defined by its computing capacity, denoted as $C_{j, k}$ and obtained by profiling the execution time of a decoder block. Correspondingly, the available memory size of a node is defined by its memory capacity, $M_{j, k}$, measured in GB. For the purpose of making model partitioning decisions, we need a metric to assess the overall service potential of each tier. To this end, we define the ``effective capacity'' of a tier based on the performance achievable within that tier, represented by the resources of its most capable node:

\begin{equation}
C_j^{\text {eff}} \triangleq \max _{k \in \mathcal{K}_j}\left\{C_{j, k}\right\}, \quad \forall j \in \mathcal{T},
\end{equation}

\begin{equation}
M_j^{\text {eff}} \triangleq \max _{k \in \mathcal{K}_j}\left\{M_{j, k}\right\}, \quad \forall j \in \mathcal{T}.
\end{equation}

Here, $C_j^{\text{eff}}$ and $M_j^{\text{eff}}$ represent the effective computing capacity and effective memory capacity of a tier, respectively. When different parts of the model are deployed across various tiers, the transmission of intermediate data introduces communication latency.

We define an inter-tier communication latency, denoted as $\tau_{j, l}$, to be the time required to transmit an activation tensor from any node in tier $j$ to any node in tier $l$. Given the linear pipeline structure of the model, the primary communication overhead occurs between adjacent tiers, we have $\tau_{j, j+1} = S_{act} / R_{j,j+1}$, where $R_{j,j+1}$ represents the effective end-to-end data rate between adjacent tiers and $S_{act}$ is the size of the activation tensor. This data rate can be expressed as $R_{j,j+1} = B_{j,j+1} \log_2(1 + \text{SINR}_{j,j+1})$, where $B_{j,j+1}$ is the allocated channel bandwidth, and $\text{SINR}_{j,j+1}$ is the Signal-to-Interference-plus-Noise Ratio at the receiving node \cite{refShannon}.

\subsection{Overall Optimization Problem}

\begin{figure}[htbp]
\centerline{\includegraphics[scale=0.3]{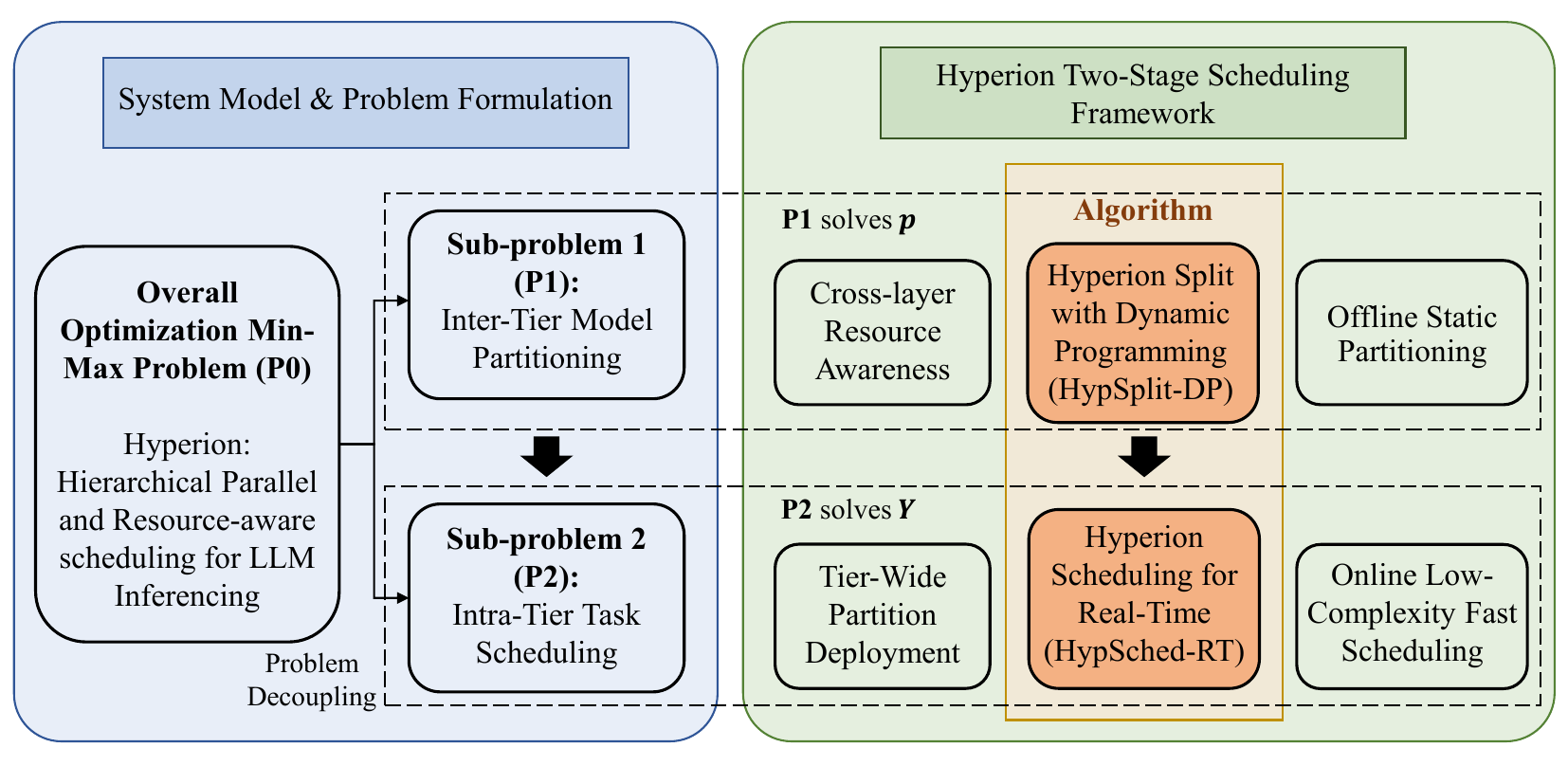}}
\caption{Framework for Problem Decomposition in Hyperion, Detailing how the Overall Optimization Problem ($\mathcal{P}_{0}$) is Decoupled into two Sub-problems, Inter-tier Partitioning ($\boldsymbol{\mathcal{P}_{1}}$) and Intra-tier Scheduling ($\boldsymbol{\mathcal{P}_{2}}$), and Solved Respectively.}
\label{fig3}
\vspace{-0.2cm}
\end{figure}

Fig. \ref{fig3} summarizes the Min–Max formulation $\mathcal{P}_{0}$ and its two subproblems. The figure outlines how this primary problem is decoupled into two manageable sub-problems: inter-tier Model Partitioning ($\mathcal{P}_{1}$) and intra-tier Task Scheduling ($\mathcal{P}_{2}$). The objective is to minimize the end-to-end latency for inference, which comprises two primary components: computing latency and communication latency.

We define two sets of decision variables for model partitioning and task scheduling. We employ a partitioning vector $\mathbf{p}=\left(p_1, p_2, \ldots, p_{T-1}\right) \in \mathbb{N}^{T-1}$ to define the partitioning scheme of the model. Herein, $p_j$ denotes the global index of the last decoder block allocated to the $j$-th tier. For convenience, we introduce the auxiliary variables $p_0 \triangleq 0$ and $p_T \triangleq N$. Consequently, the index set of the decoder blocks assigned to tier $j \in \mathcal{T}$ can be expressed as:

\begin{equation}
\mathcal{I}_j(\mathbf{p})=\left\{i \in \mathbb{N} \mid p_{j-1}<i \leq p_j\right\}.
\end{equation}

We define a set of binary variables $\mathbf{Y}=\left\{y_{j, k} \mid j \in \mathcal{T}, k \in \mathcal{K}_j\right\}$ to represent the task scheduling decisions. These variables are defined as follows:

\begin{equation}
y_{j, k}= \begin{cases}1, & \text { task of tier } j \text { scheduled to node } k, \\ 0, & \text { otherwise }.\end{cases}
\end{equation}

The total latency, denoted as $L_{\text{total}}(\mathbf{p}, \mathbf{Y})$, is composed of the computing latency $L_{\text{comp}}(\mathbf{p}, \mathbf{Y})$ and the communication latency $L_{\text{comm}}(\mathbf{p})$. For a given partitioning scheme $\mathbf{p}$, the computing workload assigned to tier $j$ is defined as $W_j(\mathbf{p}) = \sum_{i \in \mathcal{I}_j(\mathbf{p})} f_i$. If node $(j, k)$ executes this workload, its computation time is $W_j(\mathbf{p}) / C_{j, k}$. Consequently, factoring in the scheduling decision matrix $\mathbf{Y}$, the actual computation time for tier $j$ is expressed as $\sum_{k \in \mathcal{K}_j} y_{j, k} \frac{W_j(\mathbf{p})}{C_{j, k}}$. Given that the tiers operate in a pipeline fashion, the overall computing latency is dictated by the performance of the slowest stage (i.e., the bottleneck). Therefore:

\begin{equation}
L_{\text {comp }}(\mathbf{p}, \mathbf{Y})=\max _{j \in \mathcal{T}}\left\{\sum_{k \in \mathcal{K}_j} y_{j, k} \frac{\sum_{i=p_{j-1}+1}^{p_j} f_i}{C_{j, k}}\right\}.
\end{equation}

The total communication latency is the cumulative sum of the communication latency between all adjacent pipeline stages.

\begin{equation}
L_{\text {comm }}(\mathbf{p})=\sum_{j=1}^{T-1} \tau_{j, j+1}.
\end{equation}

Our goal is to minimize the total latency, which is the sum of communication and computing latency, denoted as $\boldsymbol{\mathcal{P}_{0}}$:

\begin{subequations}
\label{eq1} 
\begin{align}
    \boldsymbol{\mathcal{P}_{0}}:\min _{\mathbf{p}, \mathbf{Y}} \quad
    & \left(\max _{j \in \mathcal{T}}\left\{\sum_{k \in \mathcal{K}_j} y_{j, k} \frac{\sum_{i=p_{j-1}+1}^{p_j} f_i}{C_{j, k}}\right\}\right. \nonumber \\
    & \left. \qquad + \sum_{j=1}^{T-1} \tau_{j, j+1}\right), \label{con1} \\
    \text { s.t. } \quad 
    & 1 \leq p_1<p_2<\cdots<p_{T-1}<N, \label{con2} \\
    & \sum_{k \in \mathcal{K}_j} y_{j, k}=1, \forall j \in \mathcal{T}, \label{con3} \\
    & \sum_{i=p_{j-1}+1}^{p_j} m_i \leq \sum_{k \in \mathcal{K}_j} y_{j, k} M_{j, k}, \forall j \in \mathcal{T}, \label{con4} \\
    & p_j \in\{1,2, \ldots, N-1\}, \ \forall j \in \{1, \ldots, T-1\}, \label{con5} \\
    & y_{j, k} \in\{0,1\}, \forall j \in \mathcal{T}, \forall k \in \mathcal{K}_j. \label{con6}
\end{align}
\end{subequations}

The core objective of $\boldsymbol{\mathcal{P}_{0}}$ is to minimize the end-to-end total inference latency, which is formulated as the sum of two components. The first term in Eq. \eqref{con1} is the computing latency, which corresponds to the computation time of the longest running stage (i.e., the bottleneck) among all tiers. The second term in Eq. \eqref{con1} is the communication latency, defined as the aggregate time for transmitting intermediate activations between all adjacent tiers. This optimization problem is subject to a series of constraints. Constraint Eq. \eqref{con2} ensures the validity of the model partition vector $\mathbf{p}$, mandating that the partition must be strictly increasing and fall within a valid range, thereby partitioning the $N$ decoder blocks into $T$ non empty and contiguous subsequences. Constraint Eq. \eqref{con3} assigns each tier’s task to exactly one node. Constraint Eq. \eqref{con4} is the memory capacity constraint, which requires that the total memory footprint of all modules assigned to any given tier $j$ must not exceed the available memory capacity $M_{j,k}$ of the specific node selected by the scheduling decision $y_{j,k}$. Constraints Eq. \eqref{con5} and Eq. \eqref{con6} define the domains of the decision variables, specifying that the partition points $p_j$ are integers, while the scheduling decisions $y_{j,k}$ are binary variables.

\subsection{Problem Decoupling and Hierarchical Solution}

We compute the partition offline from fixed resource profiles. During inference, the partition is held fixed, and node selection uses the current queue and memory state. This separation avoids recomputing stage boundaries at each request arrival. Based on this, we decouple $\boldsymbol{\mathcal{P}_{0}}$ into the following two sub-problems:

\begin{itemize}
  \item \emph{Sub problem 1: Cross Hierarchy Model Partitioning ($\boldsymbol{\mathcal{P}_{1}}$)}: This is an offline planning problem that determines a contiguous partition $\mathbf{p}$ under fixed tier-level computing and memory profiles.
  
  \item \emph{Sub problem 2: Intra Hierarchy Task Scheduling ($\boldsymbol{\mathcal{P}_{2}}$)}: This is an intra-tier, dynamic decision problem. Given a fixed model partitioning scheme, when the input data for an inference request (i.e., the tokenized prompt for Tier 1, or intermediate activations for subsequent tiers) arrives at a specific hierarchy, an execution node must be selected to execute the corresponding computational task based on the real time status of all nodes within that hierarchy.
\end{itemize}

This decomposition is used to construct a computationally practical solution to $\mathcal{P}_{0}$. HypSplit-DP solves the profiled offline partitioning subproblem $\mathcal{P}_{1}$ up to the numerical tolerance $\epsilon$, whereas HypSched-RT selects the feasible node with the minimum modeled completion time for the current request at the current tier.

The core of $\boldsymbol{\mathcal{P}_{1}}$ is to determine the model partition vector, denoted as $\mathrm{p}^*$. For offline partitioning, we use fixed tier-level resource profiles. We replace the node level computing capacity $C_{j, k}$ and memory capacity $M_{j, k}$ in the original problem with the tier level effective computing capability $C_j^{\text {eff}}$ and tier level effective memory capacity $M_j^{\text {eff}}$, respectively. Consequently, the original objective function and its associated constraints are simplified, transforming it into an integer programming problem, $\boldsymbol{\mathcal{P}_{1}}$, that depends solely on the partition variable $\mathbf{p}$.

\begin{subequations}
\label{eq:problem_2} 
\begin{align}
    \boldsymbol{\mathcal{P}_{1}}:\min _{\mathbf{p}} \quad & \left(\max _{j \in \mathcal{T}}\left\{\frac{\sum_{i=p_{j-1}+1}^{p_j} f_i}{C_j^{\text {eff}}}\right\}+\sum_{j=1}^{T-1} \tau_{j, j+1}\right), \label{eq:problem_2_obj} \\
    \text { s.t. } \quad & 1 \leq p_1<p_2<\cdots<p_{T-1}<N, \label{const_a} \\
    & \sum_{i=p_{j-1}+1}^{p_j} m_i \leq M_j^{\text {eff}},  \forall j \in \mathcal{T}, \label{const_b} \\
    & p_j \in\{1,2, \ldots, N-1\},  \forall j \in\{1, \ldots, T-1\}. \label{const_c}
\end{align}
\end{subequations}

Here, $\boldsymbol{\mathcal{P}_{1}}$ is to minimize the total system latency based on the estimated performance of each tier. Constraint \eqref{const_a} defines a valid contiguous partition, constraint \eqref{const_b} requires the model chunk assigned to each tier to fit within its effective memory capacity, and constraint \eqref{const_c} defines the domain of the partition variables.

After obtaining the partition $\mathbf{p}^*$, which is computed offline, the task for each tier $j$ is determined. The subproblem $\mathcal{P}_2$ is an online scheduling problem. When the intermediate activations for an inference task arrive at tier j from tier j-1 (let the arrival time be t), the intra-tier scheduler must select an execution node for it. This decision corresponds to determining the scheduling vector for tier j, denoted as $Y_j=\{y_{j,k}\}_{k\in\mathcal{K}_j}$, which is a component of the overall decision matrix $Y$ defined in $\mathcal{P}_0$. 

Unlike the offline static partitioning problem $\mathcal{P}_{1}$, $\mathcal{P}_{2}$ is a dynamic online decision that must be made in real time as tasks arrive. A high complexity scheduling algorithm would introduce scheduling overhead, creating a new latency bottleneck that would undermine the optimization goals. Therefore, the algorithm used to solve $\mathcal{P}_{2}$ (i.e., HypSched-RT) must be lightweight and low complexity to ensure minimal overhead and meet these real time requirements.

The optimization objective here is to minimize the task's completion time at the current tier, which is composed of the node's queuing latency and the task's actual execution time. Let $Q_{j,k}(t)$ be the task queue of node (j, k) at time t, and let $T^{wait}_{j,k}(t)$ be the expected waiting time to process all tasks currently in that queue.

\begin{subequations}
\label{eq:P2} 
\begin{align}
    \boldsymbol{\mathcal{P}_{2}}: \min _{Y_j} & \sum_{k \in \mathcal{K}_j} y_{j, k}\left(T^{wait}_{j,k}(t)+\frac{\sum_{i=p_{j-1}^*+1}^{p_j^*} f_i}{C_{j, k}}\right), \\
    \text{s.t.} & \quad \sum_{\substack{k \in \mathcal{K}_j }} y_{j, k}=1, \quad y_{j, k} \in\{0,1\}, \label{eq:P2_const_a} \\
    & \sum_{i=p_{j-1}^*+1}^{p_j^*} m_i \leq M_{j, k}^{\text {avail }}(t), \label{eq:P2_const_b}
\end{align}
\end{subequations}

\section{Hyperion: A Two-Stage Hierarchical Partitioning and Scheduling Framework}

\begin{figure*}[htbp]
\centerline{\includegraphics[scale=0.5]{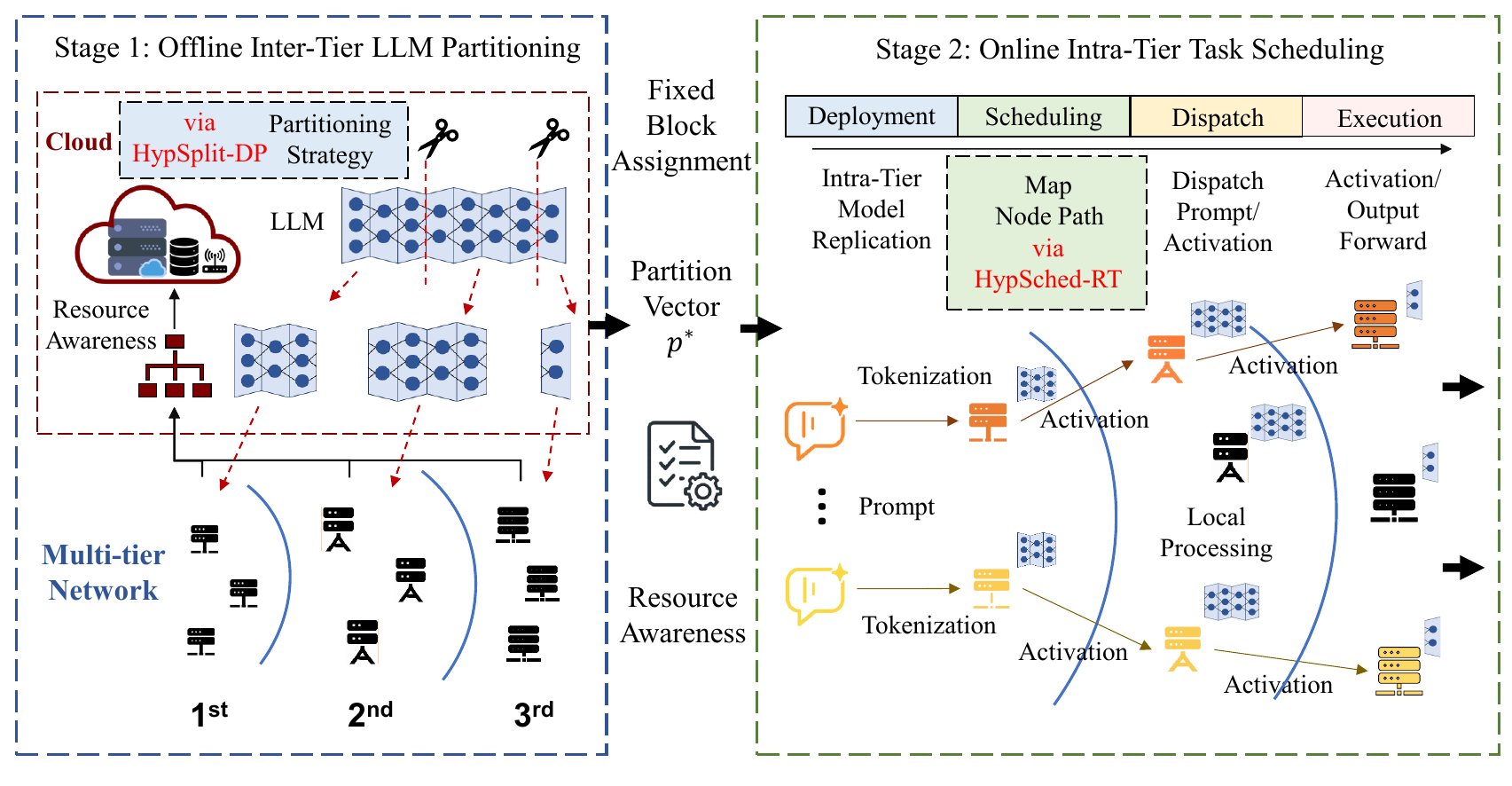}}
\caption{System Architecture and Operational Workflow of the Hyperion Framework, Illustrating the end-to-end Process from Offline LLM Partitioning in the Cloud to Online, Real Time Task Execution Across the Multi-tier Network.}
\label{fig4}
\end{figure*}

As illustrated in Fig. \ref{fig4}, the Hyperion framework operates through a two-stage process to manage LLM inference across the network. The first stage is an offline, inter-tier partitioning phase where the central cloud partitions the LLM using the tiers’ computing and memory profiles. This partitioning results in a sequential deployment of the sub models, creating a pipeline across the network tiers. The second stage consists of online, intra-tier task scheduling. Here, incoming prompts are tokenized and dispatched in real time, with the HypSched-RT algorithm dynamically mapping inference tasks to the nodes within each tier for execution, from activation input to output forwarding. This scheduling process selects a single, most suitable device from each tier's node cluster to handle its portion of the inference task.

\subsection{Stage 1: Offline Inter-Tier LLM Partitioning via HypSplit-DP}

We aim to determine the model partition vector $\mathbf{p}^*=\left(p_1^*, p_2^*, \ldots, p_{T-1}^*\right)$, which can be expressed as $\boldsymbol{\mathcal{P}_{1}}$. For notational convenience, we define the computing workload of the $j$-th tier as $L_j(\mathbf{p})=\frac{1}{C_j} \sum_{i=p_{j-1}+1}^{p_j} f_i$, where $p_0=0$ and $p_T=N$. Therefore, the objective function can be concisely expressed as $\min_{\mathbf{p}} \max_{j \in \mathcal{T}} \left\{L_j(\mathbf{p})\right\}$. For a fixed batch size, sequence length, hidden dimension, and activation data type, the activation payload at a decoder-block boundary is independent of the selected boundary. Consequently, the total inter-tier transmission time, represented by $\sum_{j=1}^{T-1} \tau_{j, j+1}$, is constant with respect to p and can be omitted from the optimization.

This problem uses a Min-Max objective to reduce the maximum profiled stage time subject to the tier-level memory constraints. A lower maximum stage time generally reduces modeled workload imbalance among the tiers and can reduce bottleneck-induced idle periods. However, the actual pipeline idle time also depends on request arrivals and the runtime execution policy. Enumerating all possible combinations of partition points would incur combinatorial search overhead. We therefore design the Hyperion Split with Dynamic Programming (HypSplit-DP) algorithm to identify a feasible partition efficiently.

We transform the original problem into a decision problem: given a target maximum latency $\tau$, does a valid partitioning scheme $p$ exist such that the computing latency of every tier does not exceed $\tau$. Specifically, we need to determine if there exists a partitioning vector $p=\left(p_1, \ldots, p_{T-1}\right)$ that satisfies the constraints of $\boldsymbol{\mathcal{P}_{1}}$ while also fulfilling the condition:

\begin{equation}
L_j(p)=\frac{\sum_{i=p_{j-1}+1}^{p_j} f_i}{C_j} \leq \tau, \quad \forall j \in \mathcal{T}.
\end{equation}

Solving this efficiently allows us to then perform a binary search for $\tau$ to find its minimum possible value, which corresponds to the solution of the original problem. We employ dynamic programming to solve the decision subproblem $P_{\text{check}}(\tau)$. We define a boolean state variable, $DP(j, n)$, which is true if and only if there exists a feasible assignment of the first $n$ decoder blocks ($B_1, \ldots, B_n$) to the first $j$ computing tiers (Tier $1, \ldots$, Tier $j$) that satisfies both the latency and memory constraints.

\begin{myDef}

The state $DP(j, n)$ is true if and only if there exists a partition point $k$ (where $j-1 \leq k < n$) such that the following two conditions are satisfied:
\begin{itemize}
\item The first $k$ modules can be feasibly partitioned among the first $j-1$ tiers; that is, $DP(j-1, k)$ is true.
\item The modules from $k+1$ to $n$ are assigned as a single group to the $j$-th tier, and this assignment satisfies the latency and memory constraints of the given decision problem.
\end{itemize}

\end{myDef}

This logical relationship can be described by the following state transition:

\begin{equation}
\begin{split}
DP(j, n) = \bigvee_{k=j-1}^{n-1} \bigg( & DP(j-1, k) \wedge \left[\frac{1}{C_j} \sum_{i=k+1}^n f_i \leq \tau\right] \wedge \\
& \left[\sum_{i=k+1}^n m_i \leq M_j\right] \bigg),
\end{split}
\end{equation}
in which $\vee$ denotes the logical OR operation, $\wedge$ denotes the logical AND operation, and $[\cdot]$ is the Iverson bracket, which evaluates to 1 if the condition inside is true, and 0 otherwise. To improve computing efficiency, we precompute the prefix sums of $f_i$ and $m_i$:

\begin{equation}
S_f(n)=\sum_{i=1}^n f_i,
\end{equation}

\begin{equation}
S_m(n)=\sum_{i=1}^n m_i.
\end{equation}

Hence, the interval sums can be computed efficiently: $\sum_{i=k+1}^n f_i=S_f(n)-S_f(k)$ and $\sum_{i=k+1}^n m_i=S_m(n)-S_m(k)$. For the base case, at the first tier ($j=1$), the state $DP(1, n)$ is true if and only if placing all of the first $n$ modules on the first tier satisfies the constraints:

\begin{equation}
D P(1, n)=\left[\frac{S_f(n)}{C_1} \leq \tau\right] \wedge\left[S_m(n) \leq M_1\right].
\end{equation}

The solution to the entire decision problem $P_{\text{check}}(\tau)$ is determined by the value of $DP(T, N)$. If $DP(T, N)$ is true, it indicates that a feasible scheme exists with a latency not exceeding $\tau$; otherwise, no such scheme exists.

\begin{algorithm}[t]
\caption{HypSplit-DP for Inter-tier Model Partitioning}
\label{alg:model_partitioning_revised}
\begin{algorithmic}[1]
\STATE \textbf{Input:} Number of blocks \emph{$N$}, tiers \emph{$T$}, block costs \emph{$\{f_i\}, \{m_i\}$}, tier properties \emph{$\{C_j\}, \{M_j\}$}, and precision \emph{$\epsilon$}.
\STATE \textbf{Output:} Partition vector \emph{$\mathbf{p}^*$} and minimized maximum latency \emph{$\tau^*$}.
\vspace{0.5em}
\STATE Compute prefix sums: \emph{$S_f(n) \gets \sum_{i=1}^{n} f_i$} and \emph{$S_m(n) \gets \sum_{i=1}^{n} m_i$}.
\STATE Initialize binary search bounds \emph{$\tau_{\text{low}}$} and \emph{$\tau_{\text{high}}$}.
\STATE Initialize \emph{$\mathbf{p}^* \gets \text{null}$}, \emph{$\tau^* \gets \tau_{\text{high}}$}.
\vspace{0.5em}
\WHILE{\emph{$(\tau_{\text{high}} - \tau_{\text{low}}) > \epsilon$}}
    \STATE Set current target latency \emph{$\tau_{\text{mid}} \gets (\tau_{\text{low}} + \tau_{\text{high}}) / 2$}.
    \STATE Initialize DP table \emph{$DP(j,n)$} to \emph{false} and predecessor table \emph{$P[j][n]$}.
    \STATE Set \emph{$DP[0][0] \gets \text{true}$}.
    \vspace{0.5em}
    \FOR{\emph{$j \gets 1$ to $T$}}
        \FOR{\emph{$n \gets j$ to $N$}}
            \FOR{\emph{$k \gets j-1$ to $n-1$}}
                \IF{\emph{$DP[j-1][k]$} is \emph{true}}
                    \STATE Calculate computing workload \emph{$L_j \gets  (S_f(n) - S_f(k)) / C_j$}.
                    \STATE Calculate memory usage \emph{$U_j \gets S_m(n) - S_m(k)$}.
                    \IF{\emph{$L_j \leq \tau_{\text{mid}}$} and \emph{$U_j \leq M_j$}}
                        \STATE Set \emph{$DP(j,n) \gets \text{true}$}.
                        \STATE Record predecessor \emph{$P[j][n] \gets k$}.
                        \STATE \textbf{Break}.
                    \ENDIF
                \ENDIF
            \ENDFOR
        \ENDFOR
    \ENDFOR
    \vspace{0.5em}
    \IF{\emph{$DP[T][N]$} is \emph{true}}
        \STATE A feasible partition found, update latency \emph{$\tau^* \gets \tau_{\text{mid}}$}.
        \STATE Adjust search upper bound \emph{$\tau_{\text{high}} \gets \tau_{\text{mid}}$}.
        \STATE Reconstruct partition vector \emph{$\mathbf{p}^*$} by backtracking through \emph{$P$}.
    \ELSE
        \STATE No feasible partition found, adjust search lower bound \emph{$\tau_{\text{low}} \gets \tau_{\text{mid}}$}.
    \ENDIF
\ENDWHILE
\vspace{0.5em}
\STATE \textbf{Return} \emph{$\mathbf{p}^*$} and \emph{$\tau^*$}.
\end{algorithmic}
\end{algorithm}

As illustrated in Algorithm 1, we first determine the initial lower and upper bounds of the latency, $\tau_{\text{low}}$ and $\tau_{\text{high}}$. In each iteration, we select a midpoint value $\tau_{\text{mid}}$ and solve a feasibility subproblem, $P_{\text{check}}(\tau_{\text{mid}})$, to ascertain whether the latency $\tau_{\text{mid}}$ is achievable. This feasibility subproblem is efficiently solved using dynamic programming (DP).

Based on the final state of the DP table, $DP(T, N)$, we adjust the boundaries for the binary search. If the state is true, it implies that a better solution may exist; thus, we update the upper bound to $\tau_{\text{high}} = \tau_{\text{mid}}$ and record the currently feasible partitioning scheme. If it is false, it indicates that the current latency is too stringent and must be relaxed, so we update the lower bound to $\tau_{\text{low}} = \tau_{\text{mid}}$. This process iterates until the search interval is smaller than a predefined precision threshold $\epsilon$. Finally, the algorithm converges to the objective value, $\tau_{\text{high}}$, and the task partitioning vector, $p^*$, is retrieved by backtracking through the last successful DP table. The overall complexity of this algorithm is $O(T \cdot N^2 \cdot \log(\frac{\tau_{\text{high}} - \tau_{\text{low}}}{\epsilon}))$.

HypSplit-DP converges to the solution of subproblem $\mathcal{P}_{1}$ within the tolerance $\epsilon$. In each search iteration, a DP check, $P_{check}(\tau)$, serves as an exhaustive feasibility checker. Its state $DP(j,n)$ and transition function systematically exhaust all valid combinations for partitioning $N$ blocks into $T$ tiers by iterating through all possible preceding split points $k$.

\subsection{Stage 2: Online intra-tier Task Scheduling via HypSched-RT}

Once the model partition $p^*$ is determined, the task assigned to tier $j$ becomes fixed, with a total computing workload of $F_j^* = \sum_{i=p_{j-1}^*+1}^{p_j^*} f_i$. When an inference task arrives at tier $j$ at time $t$, the scheduler must select a node $k$ from the set of nodes at that tier, $\mathcal{K}_j$, to execute the task. This selection process is the online scheduling subproblem $\mathcal{P}_{2}$, which we rewrite in node-selection form as follows:

\begin{subequations}
\label{eq:k_star_selection} 
\begin{align}
    k^* = & \arg \min _{k \in \mathcal{K}_j}\left(T^{wait}_{j,k}(t)+\frac{\sum_{i=p_{j-1}^*+1}^{p_j^*} f_i}{C_{j, k}}\right) \label{eq:k_star_obj}, \\
    \text{s.t.} \quad & \text{Node } k \text{ is available at time } t, \label{eq:k_star_const_a} \\
    & \sum_{i=p_{j-1}^*+1}^{p_j^*} m_i \leq M_{j, k}^{\text {avail }}(t), \label{eq:k_star_const_b}
\end{align}
\end{subequations}

We propose the Hyperion Scheduling for Real-Time (HypSched-RT) algorithm to handle the online challenge of arriving inference tasks. HypSched-RT enables intra-tier task parallelism by scheduling inference tasks across a tier's nodes ($K_j$). We maintain a task queue $Q_{j, k}$, for each node $(j, k)$, which enables the estimation of the waiting time $T^{wait}_{j,k}(t)$. The task queue $Q_{j,k}$ is essential because individual edge nodes, such as Jetson devices, possess limited parallel inference capabilities. When a new task arrives, we assume it is placed at the end of this queue, following a First In First Out (FIFO) strategy. At a given time $t$, the expected waiting time for node $(j, k)$ is the sum of the remaining processing times for all tasks currently executing and those waiting in its queue. Let $q_{j, k}^{run}(t)$ be the task being executed on node $(j, k)$ at time $t$, and let $Q_{j, k}^{\text{wait}}(t)$ be the set of tasks waiting in the queue of node $(j, k)$ at time $t$. Let $F_q$ denote the total computing workload of a task $q$, and let $F_{q, rem}(t)$ be the remaining computing workload of task $q$ at time $t$. The expected waiting time can be calculated as:

\begin{equation}
T^{wait}_{j,k}(t)=\frac{F_{q_{j, k}^{run}(t),\text { rem }} (t)+\sum_{q^{\prime} \in Q_{j, k}^{\text {wait }}(t)} F_{q^{\prime}}}{C_{j, k}}.
\end{equation}

If no task is currently being executed, then $F_{q_{j, k}^{\text{run}}, \mathrm{rem}}(t)=0$. For a new task with a computing workload of $F_j^*$ arriving at tier $j$ at time $t$, the scheduling decision $k^*$ is to select the node that yields the earliest completion time. The completion time of the task on node $k$, denoted as $T_{\text {complete }}(j, k, t)$, is:

\begin{equation}
\begin{split}
T_{\text{complete}}(j, k, t) ={} & t + T^{wait}_{j,k}(t) + \frac{F_j^*}{C_{j, k}}.
\end{split}
\end{equation}

Since the current time $t$ is a constant, it can be dropped from the objective. Let $\mathcal{K}_j^{\text{avail}}(t)$ denote the set of all nodes in tier $j$ that satisfy the availability and real time memory constraints at time $t$. The scheduling decision, $k^*$, is given by:

\begin{equation}
\begin{split}
k^* = \arg \min _{k \in \mathcal{K}_j^{\text{avail}}(t)} \bigg\{ & \frac{F_{q_{j, k}^{\text{run}}(t), \text{rem}}(t) + \sum_{q' \in Q_{j, k}^{\text{wait}}(t)} F_{q'}}{C_{j, k}} \\
& + \frac{F_j^*}{C_{j, k}} \bigg\}.
\end{split}
\end{equation}

\begin{algorithm}[t]
\caption{HypSched-RT for Intra-tier Scheduling}
\label{alg:intra_tier_scheduling_revised}
\begin{algorithmic}[1]
\STATE \textbf{Input:} A new task with workload \emph{$F_j^*$}; the set of nodes \emph{$K_j$} in tier \emph{$j$}; the real time state for each node \emph{$k \in K_j$} (including \emph{$C_{j,k}$}, \emph{$M_{j,k}^{\text{avail}}(t)$}, \emph{$F_{q_{j,k}^{\text{run}}(t), \text{rem}}(t)$}, and \emph{$Q_{j,k}^{\text{wait}}(t)$}); required memory \emph{$m^{\text{act}}$}.
\STATE \textbf{Output:} The node for execution, \emph{$k^*$}.

\STATE Initialize minimum completion cost \emph{$\text{min\_cost} \gets \infty$}, and node \emph{$k^* \gets \text{null}$}.

\FOR{each node \emph{$k \in K_j$}}
    \IF{node \emph{$k$} is available and satisfies memory constraint \emph{$M_{j,k}^{\text{avail}}(t) \ge m^{\text{act}}$}}
        \STATE Calculate the total workload of tasks already in its queue: \emph{$F_{k, \text{queued}} \gets F_{q_{j,k}^{\text{run}}(t), \text{rem}}(t) + \sum_{q' \in Q_{j,k}^{\text{wait}}(t)} F_{q'}$}.
        
        \STATE Compute the completion cost for the new task on node \emph{$k$}: \emph{$\text{cost}_k \gets (F_{k, \text{queued}} + F_j^*) / C_{j,k}$}.
        
        \IF{\emph{$\text{cost}_k < \text{min\_cost}$}}
            \STATE Update the minimum cost: \emph{$\text{min\_cost} \gets \text{cost}_k$}.
            \STATE Designate node \emph{$k$} as the current node: \emph{$k^* \gets k$}.
        \ENDIF
    \ENDIF
\ENDFOR

\STATE Return the determined node \emph{$k^*$}.
\end{algorithmic}
\end{algorithm}

As illustrated in Algorithm 2, when a task arrives at any tier $j$ of the network at time $t$, the scheduler immediately gathers real-time status from all nodes $k \in K_j$ within that tier. This information includes each node's available memory, $M_{j,k}^{\text{avail}}(t)$, and the current load of its task queue, encompassing the remaining workload of the running task, $F_{q_{j,k}^{\mathrm{run}}(t),\mathrm{rem}}(t)$, and the tasks in its waiting queue, $Q_{j,k}^{\text{wait}}(t)$. Subsequently, the algorithm filters these nodes based on availability and memory constraints to establish a set of candidate nodes, $K_j^{\text{avail}}(t)$. Finally, by calculating the expected completion time for the task on each candidate node $k$ as $T_{\text{complete}}(j,k,t) = t + T^{wait}_{j,k}(t) + F_j^* / C_{j,k}$, the scheduler assigns the task to the node $k^*$ that offers the minimum expected completion time. Given the fixed partition and the currently feasible-node set, HypSched-RT scans the nodes once and selects the node with the minimum completion time for the current request. If each node maintains its aggregate remaining queued workload as part of its real-time state, the required computation for each node is constant time, yielding an $O(K_j)$ per-decision complexity.

\section{Experiment and Model Evaluation}

\subsection{Experimental Setup and Baselines}

To evaluate the performance of the Hyperion framework, we designed the following network. The testbed has three tiers of NVIDIA Jetson devices. Following the pipeline parallelism model, the inference tasks flow sequentially from Tier 1 to Tier 3. The specific configurations for each tier, including the device type, quantity, computing capacity, and memory capacity, are detailed in Table \ref{tab1}. We evaluate Hyperion on LLaMA 3-8B and Phi-3-medium models, with their architectural differences detailed in Table \ref{tab1}. In our evaluation, each ``task" refers to a single, individual prompt request. Unless otherwise stated, the output length is 256 tokens per request.

\begin{table}[t]
\centering
\caption{Specifications of the Multi-tier Network and LLM Architectures}
\label{tab1}
\resizebox{\columnwidth}{!}{%
\begin{tabular}{l l c c c}
\hline
\multicolumn{5}{c}{\textbf{Hardware Specifications}} \\
\hline
\textbf{Tier} & \textbf{Device} & \textbf{Qty.} & \textbf{TOPS} & \textbf{Mem (GB)} \\
\hline
Tier 1 & J. Orin Nano & 3 & 67 & 8 \\
Tier 2 & J. Orin NX & 3 & 157 & 16 \\
Tier 3 & J. AGX Orin & 2 & 200 & 32 \\
\hline
\multicolumn{5}{c}{} \\[-1.0ex]
\hline
\multicolumn{5}{c}{\textbf{Model Architectures}} \\
\hline
\textbf{Model} & \textbf{Params} & \textbf{Blocks} & \textbf{Hid. Dim.} & \textbf{Attention} \\
\hline
LLaMA 3-8B & 8B & 32 & 4096 & GQA \\
Phi-3-medium & 14B & 40 & 5120 & GQA \\
\hline
\end{tabular}%
}
\end{table}


We use Linux TC to impose two controlled effective-link-rate settings of 1 Gbps and 100 Mbps \cite{refJupiter}. For HypSplit-DP, the binary search terminates when $\tau_{\mathrm{high}}-\tau_{\mathrm{low}}<10^{-3}$ s. For all compared methods, the LLM is partitioned at the decoder-block granularity, and only contiguous block assignments are considered. All methods use the same model implementation, block workload and memory profiles, three-tier hardware configuration, available nodes, link-rate settings, and request-arrival sequences. The same device-memory constraints are enforced when constructing a feasible partition.

After model deployment, the same inference execution environment and FIFO request queues are used for all methods. End-to-end latency is measured from the arrival of a prompt until the completion of its generated output and includes computation, queueing, and inter-tier activation transmission. The methods differ only in their partitioning and node-selection rules described below.

\begin{itemize}

\item \emph{Memory-Aware HEFT} (\textbf{HEFT}): This baseline uses a memory-aware greedy procedure to group the ordered LLM blocks into contiguous stages. Starting from the first block, blocks are assigned sequentially while respecting the memory capacity of the corresponding tier. After the stage boundaries are fixed, HEFT \cite{ref56} is used for task scheduling. Following its two-phase rule, tasks are prioritized according to their upward ranks and are assigned to the feasible node that provides the earliest finish time. The finish-time estimate uses the task computation cost and the node's available time.

\item \emph{GPipe-Style Partitioning with GNN-Based Scheduling} (\textbf{GPipe}): Following the cost-balancing principle of GPipe \cite{ref33}, this baseline divides the ordered Transformer blocks into contiguous stages so that the estimated workloads of the pipeline stages are balanced. Partitions that violate the memory capacity of a tier are excluded. Once the stage boundaries are fixed, the GNN scheduler in \cite{ref17} selects an execution node within each tier based on the profiled task demand and node-resource information. The resulting partition remains fixed throughout each experimental run.

\item \emph{PipeDream-Style Dynamic Programming with Weighted Round-Robin} (\textbf{PD-WRR}): Following the profiled stage-construction principle of PipeDream \cite{ref34}, this baseline uses dynamic programming to partition the ordered blocks into contiguous pipeline stages. The partitioning procedure balances the estimated stage workloads subject to the same tier-memory constraints used by the other methods. After the partition is fixed, requests are dispatched among the nodes of tier $j$ using weighted round-robin (WRR) \cite{ref57}. The weight assigned to node $k$ is proportional to its computing capacity:

\begin{equation}
P(k)=\frac{C_{j,k}}{\sum_{k'\in\mathcal{K}_j} C_{j,k'}}.
\end{equation}

Unlike real-time queue-based selection, this dispatch rule uses fixed capacity-derived weights and does not update them according to instantaneous queue lengths.

\item \emph{Compute-Proportional Partitioning with Weighted Join-the-Shortest-Queue} (\textbf{CP-JSQ}): This baseline combines a capacity-based static partitioner with a queue-aware dispatch policy. Compute-Proportional Partitioning (CPP) sequentially assigns contiguous LLM blocks according to the relative computing capacities of the tiers, subject to the same memory-feasibility constraints. During inference, Weighted Join-the-Shortest-Queue (W-JSQ) \cite{ref58} selects the node with the smallest capacity-normalized waiting-queue length:

\begin{equation}
k^*=\arg\min_{k\in\mathcal{K}_j}
\frac{|Q_{j,k}^{\mathrm{wait}}(t)|}{C_{j,k}}.
\end{equation}

\item \emph{Hierarchical Parallel Execution and resource-aware scheduling for LLM Inference on multi-tier Networks} (\textbf{Hyperion}): Hyperion uses HypSplit-DP for offline inter-tier model partitioning and HypSched-RT for online intra-tier request scheduling. HypSplit-DP selects memory-feasible contiguous stage boundaries by minimizing the maximum estimated stage time. HypSched-RT selects a feasible node according to its estimated completion time, including the remaining workload of the running and queued requests.

\end{itemize}

\subsection{Comparative Analysis of End-to-End Latency}

\begin{figure}[t]
    \centering
    \begin{minipage}[b]{0.85\linewidth}
        \centering
        \includegraphics[width=\linewidth]{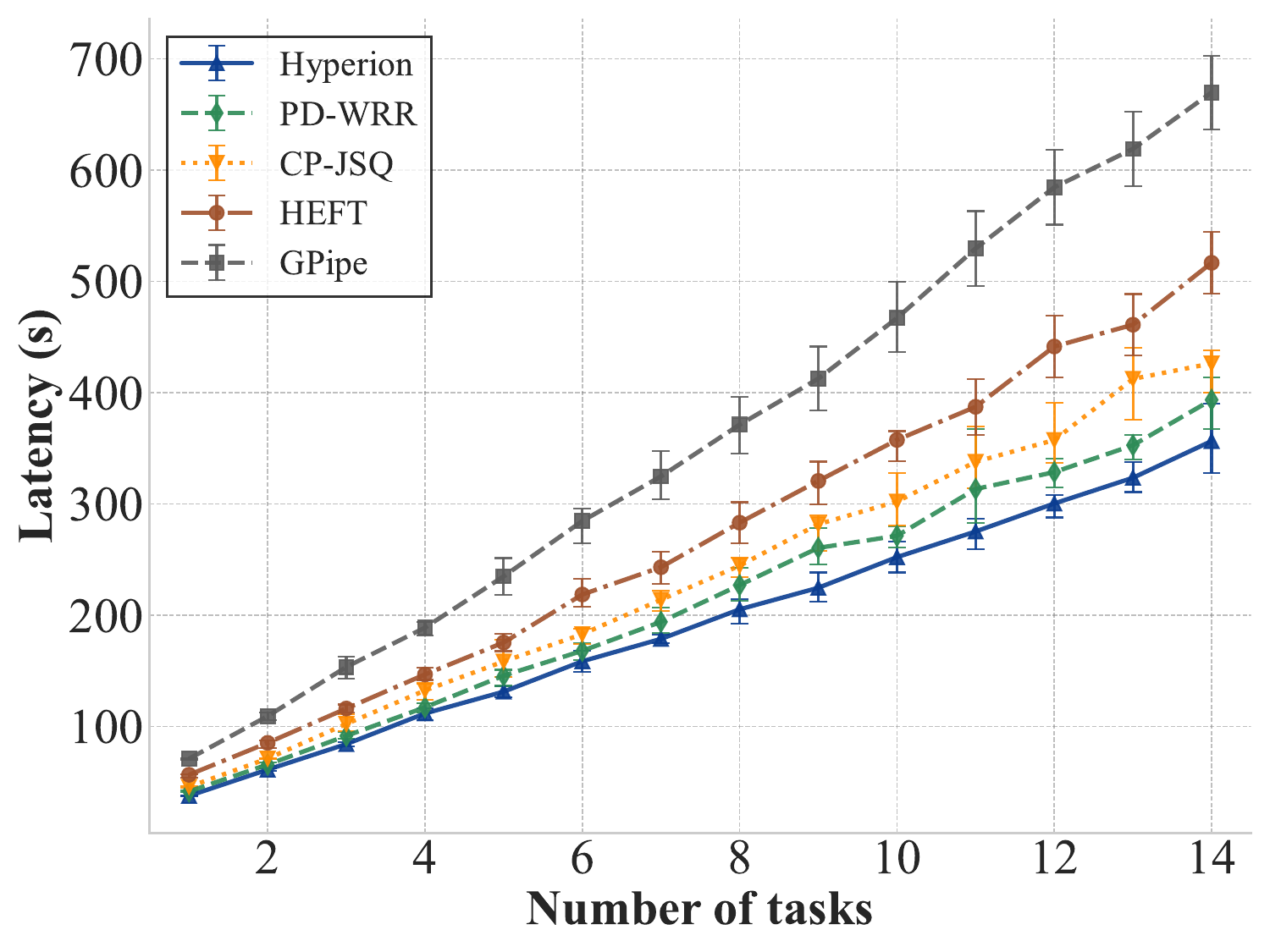}
        \subcaption{1 Gbps}
    \end{minipage}
    \vskip\baselineskip
    \begin{minipage}[b]{0.85\linewidth}
        \centering
        \includegraphics[width=\linewidth]{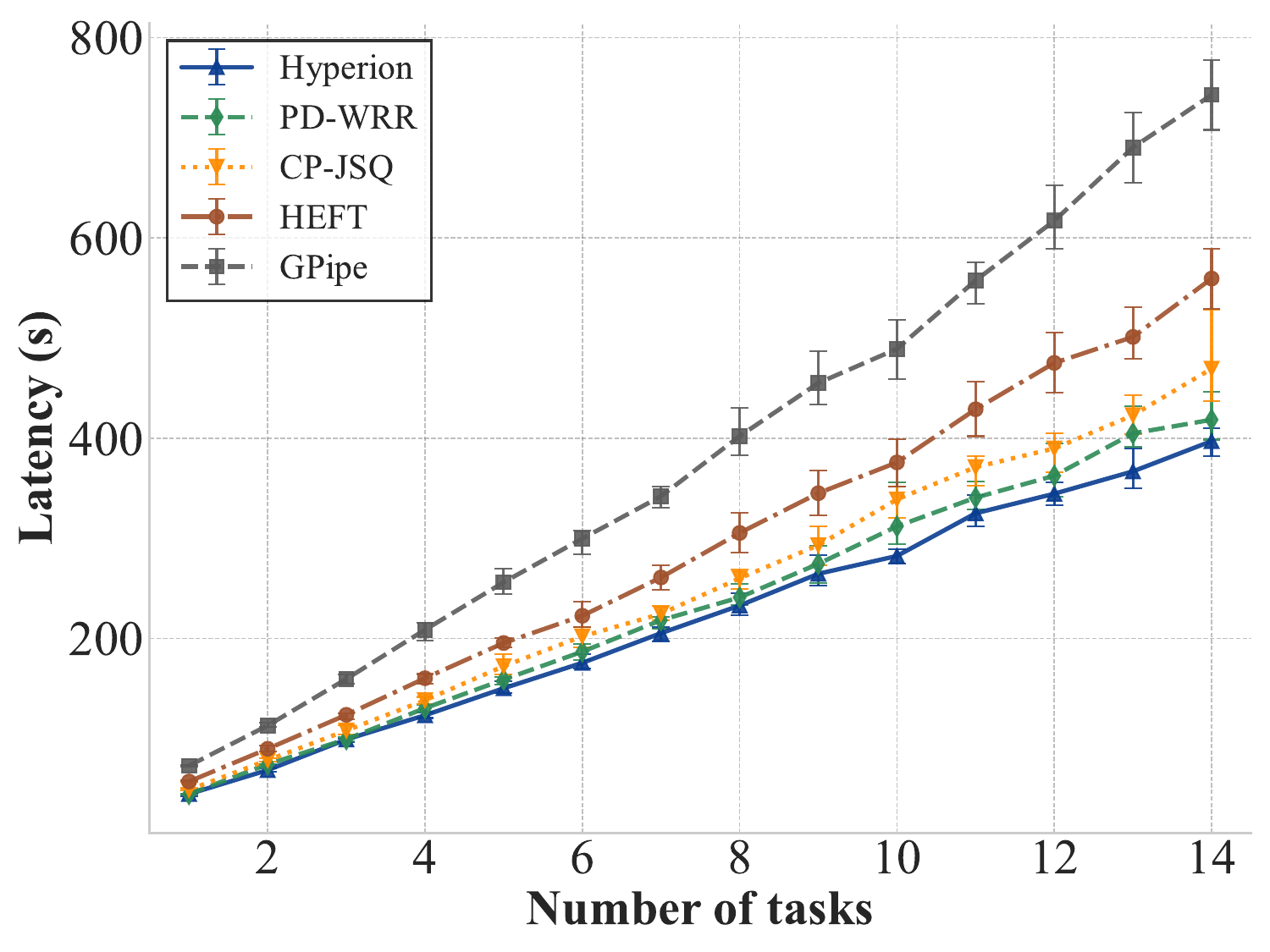}
        \subcaption{100 Mbps}
    \end{minipage}
    \caption{Comparative Analysis of End-to-End Latency for Hyperion using Llama 3 Model under Increasing Task Loads at (a) 1 Gbps and (b) 100 Mbps Bandwidths.} \label{test1}
\end{figure}

We conduct experiments within the three-tier architecture, and the results are shown in Fig. \ref{test1}. Using Llama 3 with 1 Gbps in Fig. \ref{test1}(a), GPipe and HEFT exhibit higher latency than Hyperion, while PD-WRR and CP-JSQ provide closer comparisons. At 14 tasks, the measured latencies of GPipe, HEFT, CP-JSQ, PD-WRR, and Hyperion are approximately 669, 516, 425, 393, and 355s, respectively. Hyperion therefore records 9.5\%, 16.4\%, 31.1\%, and 46.9\% lower latency than PD-WRR, CP-JSQ, HEFT, and GPipe, respectively. WRR reduces to round-robin dispatch that ignores queue state; the gap between PD-WRR and Hyperion therefore reflects both the different partitions and the queue-aware dispatch of HypSched-RT.

In Fig. \ref{test1}(b), we test at a constrained bandwidth of 100 Mbps. As the number of tasks increases from 1 to 14, the latencies for GPipe and HEFT increase to approximately 743s and 559s, respectively. PD-WRR and CP-JSQ yield latencies of approximately 418s and 465s. In contrast, Hyperion records a latency of approximately 401s. Hyperion therefore maintains lower latency than the four baselines as the number of tasks increases.

\begin{figure}[t]
    \centering
    \includegraphics[width=0.8\linewidth]{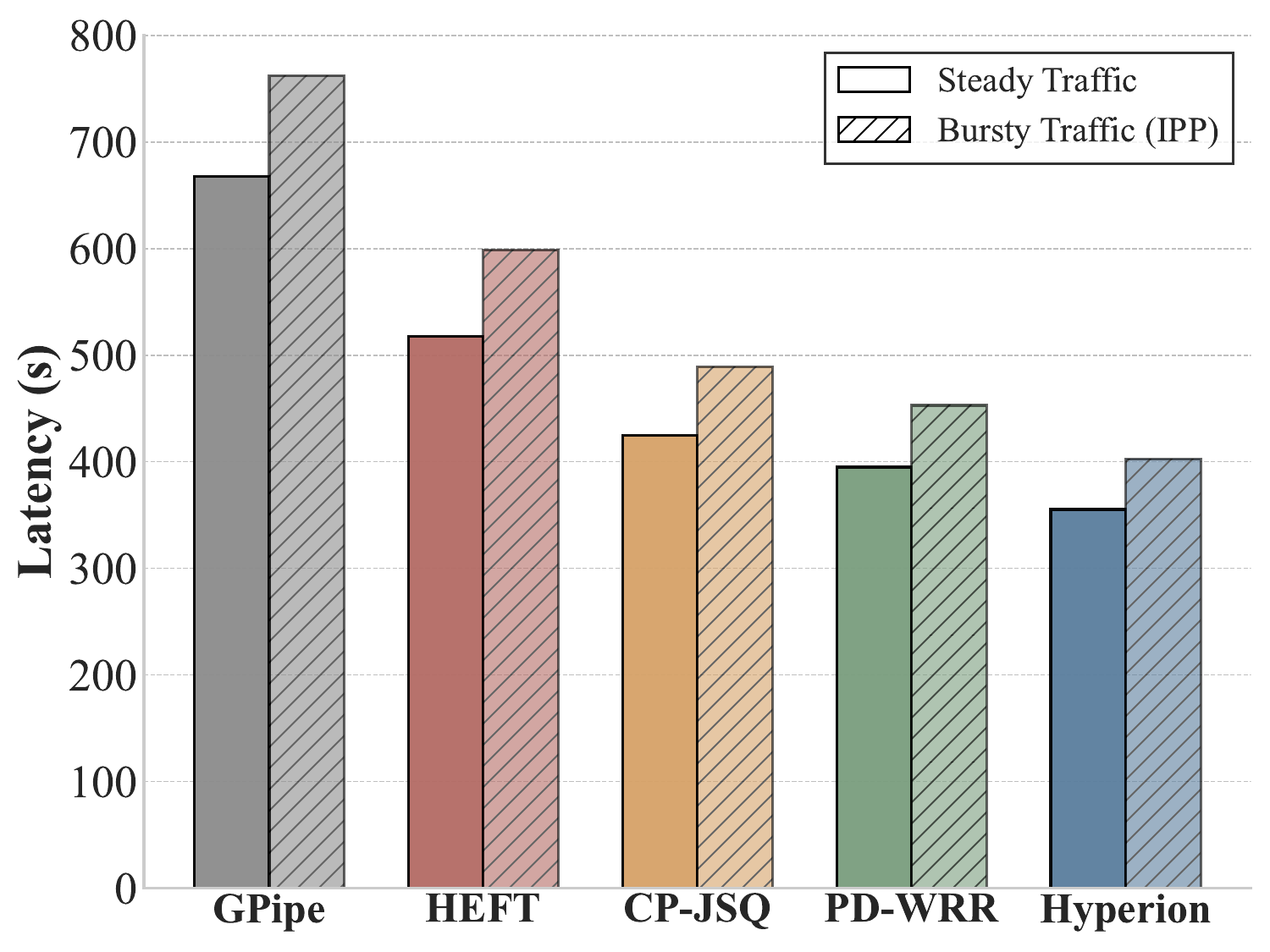} 
    \caption{End-to-End Latency Comparison under Steady and Bursty (IPP) Traffic Conditions.} \label{test1.5}
\end{figure}

Production LLM-serving traces exhibit substantial temporal burstiness \cite{refBurstGPT}. To examine performance under bursty arrivals, we use an interrupted Poisson process (IPP), a classical ON/OFF-modulated Poisson model \cite{refIPP}, as a controlled synthetic stress test. The source is in the ON state for a stationary fraction of 20\%, with $\lambda_{\mathrm{ON}}=1.0$ task/s, and in the OFF state for a stationary fraction of 80\%, with $\lambda_{\mathrm{OFF}}=0$, yielding a long-run mean arrival rate of 0.2 task/s. Fig. \ref{test1.5} compares the methods under steady and IPP bursty arrivals at 1 Gbps with 14 tasks. Under bursty traffic, the latencies of GPipe and HEFT increase from approximately 668s and 518s to 762s and 599s, corresponding to increases of 14.1\% and 15.6\%, respectively. The latencies of PD-WRR and CP-JSQ increase to approximately 452s and 488s. 

Under the same IPP setting, Hyperion's latency increases by 13.2\%, from approximately 355s to 402s. The increase for Hyperion is comparable to that of the baselines, indicating that its latency advantage is retained under bursty arrivals.

\begin{figure}[t]
    \centering
    \begin{minipage}[b]{0.85\linewidth}
        \centering
        \includegraphics[width=\linewidth]{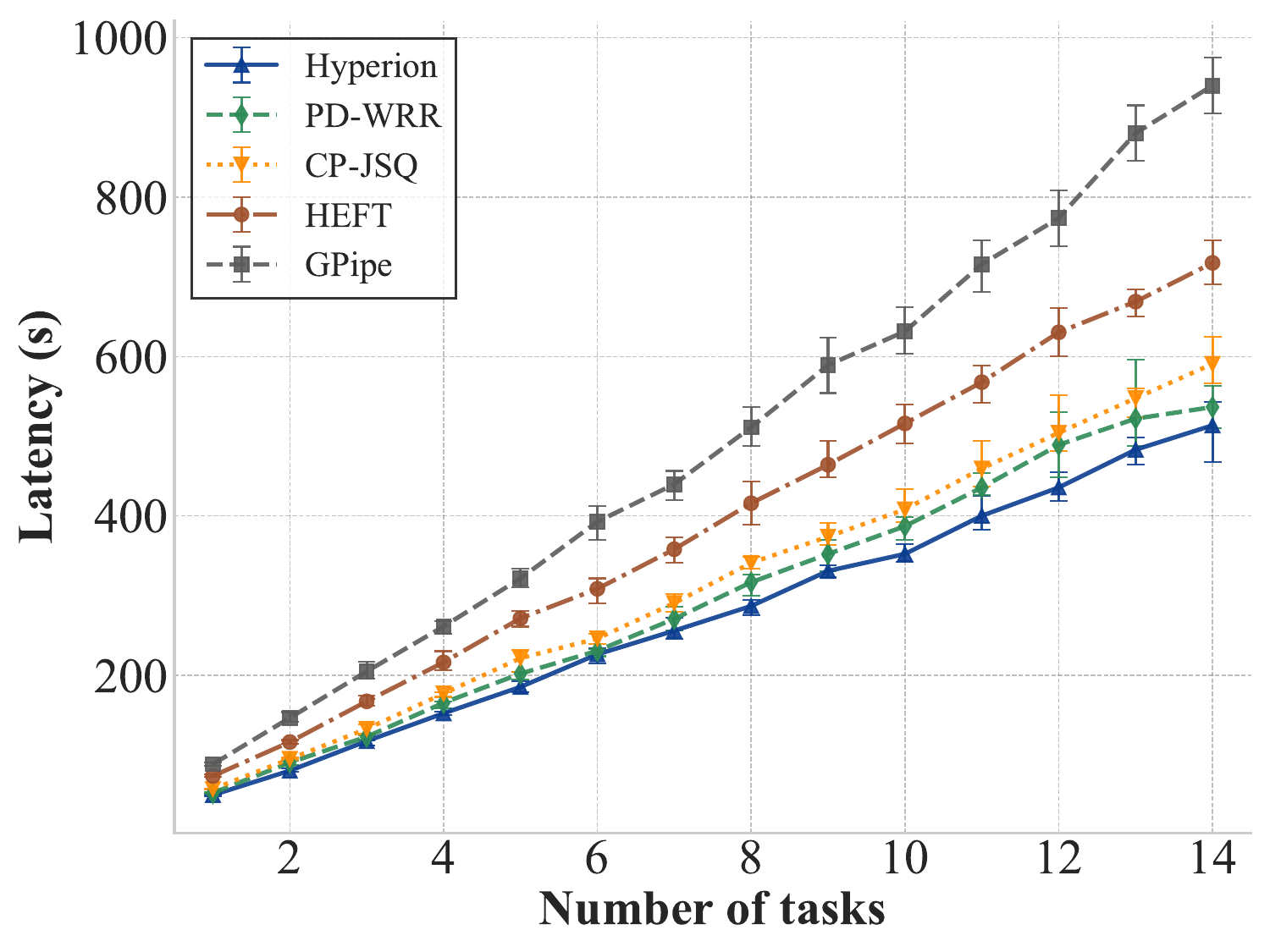}
        \subcaption{1 Gbps}
    \end{minipage}
    \vskip\baselineskip
    \begin{minipage}[b]{0.85\linewidth}
        \centering
        \includegraphics[width=\linewidth]{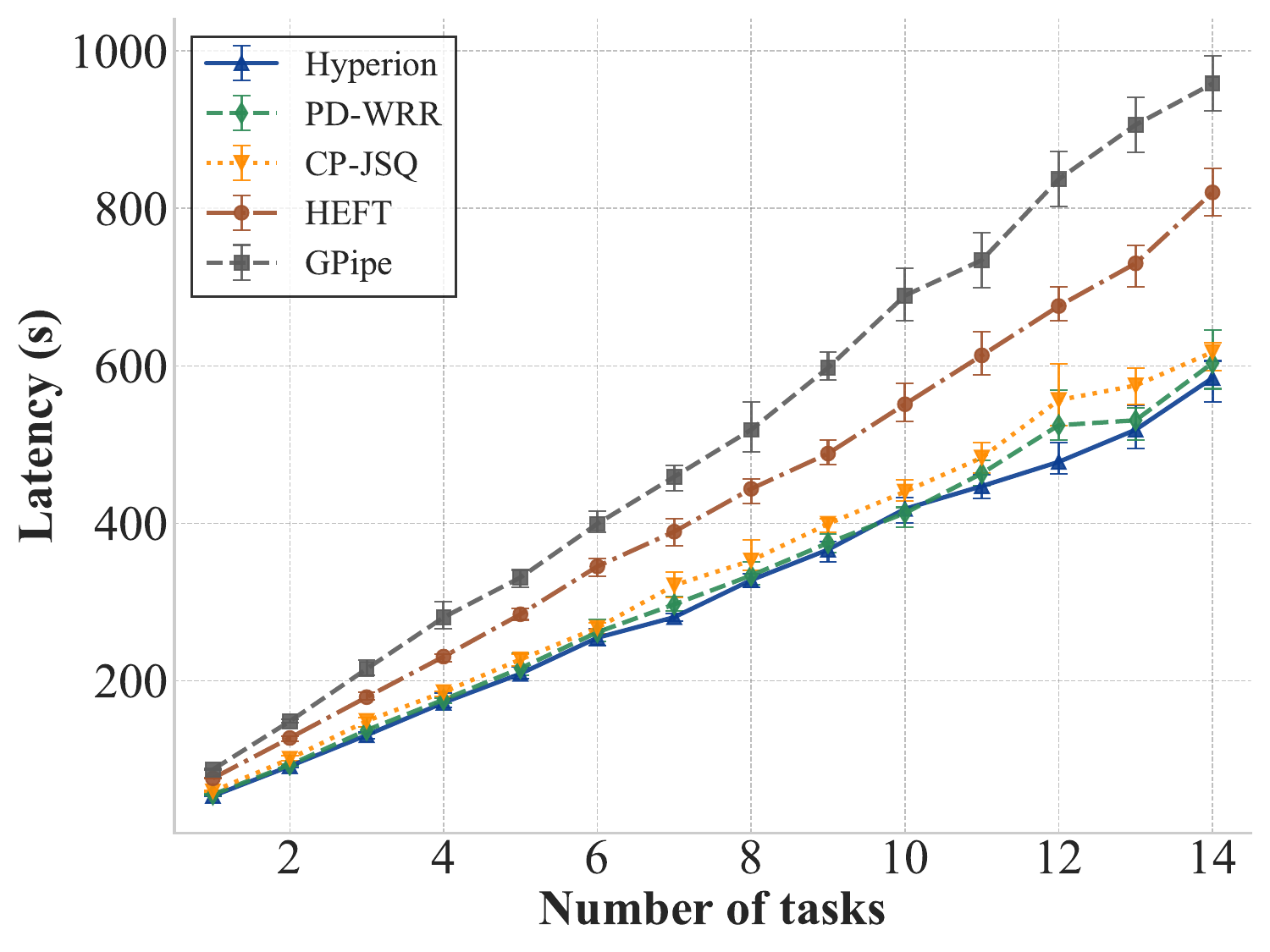}
        \subcaption{100 Mbps}
    \end{minipage}
    \caption{Comparative Analysis of End-to-End Latency for Hyperion using the Phi-3-medium Model under Increasing Task Loads at (a) 1 Gbps and (b) 100 Mbps Bandwidths.} \label{test2}
\end{figure}

We also evaluate Hyperion with the larger Phi-3-medium model. In Fig. \ref{test2}(a) (1 Gbps), Hyperion has the lowest latency over the tested task counts. When the number of tasks reaches 10, the latency for Hyperion is approximately 352s. The baselines all require more time: GPipe and HEFT take about 632s and 516s, respectively. PD-WRR requires roughly 387s, exhibiting a 9.9\% higher latency than Hyperion. Similarly, CP-JSQ yields a latency of roughly 408s (15.9\% higher than Hyperion).

At 100 Mbps, as shown in Fig. \ref{test2}(b), Hyperion also records lower latency than the four baselines. The gap widens with the task count.

\begin{table*}[t]
\centering
\caption{Latency and resource utilization for Llama 3 and Phi-3-medium at two link rates.}
\label{tab2}
\begin{tabular}{@{}lllccc@{}}
\toprule
& & & \multicolumn{3}{c}{\textbf{Jetson Device}} \\ 
\cmidrule(l){4-6}
\textbf{Model} & \textbf{Condition} & \textbf{Metric} & \textbf{AGX Orin} & \textbf{Orin NX} & \textbf{Orin Nano} \\
\midrule
\multirow{6}{*}{LLaMA 3-8B} & \multirow{3}{*}{1 Gbps Bandwidth} & Avg. GPU Util. (\%) & 45.9\% & 47.3\% & 52.1\% \\
& & Avg. Mem. Util. (\%) & 42.7\% & 45.2\% & 53.1\% \\
\cmidrule(l){3-6}
& & End-to-End Latency (s) & \multicolumn{3}{c}{24.8} \\
\cmidrule(l){2-6} 
& \multirow{3}{*}{100 Mbps Bandwidth} & Avg. GPU Util. (\%) & 43.3\% & 45.4\% & 50.1\% \\
& & Avg. Mem. Util. (\%) & 41.4\% & 44.8\% & 53.2\%\\
\cmidrule(l){3-6}
& & End-to-End Latency (s) & \multicolumn{3}{c}{27.2} \\
\midrule 
\multirow{6}{*}{Phi-3-medium} & \multirow{3}{*}{1 Gbps Bandwidth} & Avg. GPU Util. (\%) & 53.6\% & 55.4\% & 59.3\% \\
& & Avg. Mem. Util. (\%) & 69.2\% & 78.3\% & 83.1\% \\
\cmidrule(l){3-6}
& & End-to-End Latency (s) & \multicolumn{3}{c}{39.7} \\
\cmidrule(l){2-6} 
& \multirow{3}{*}{100 Mbps Bandwidth} & Avg. GPU Util. (\%) & 49.7\% & 51.8\% & 55.6\% \\
& & Avg. Mem. Util. (\%) & 69.2\% & 78.4\% & 83.3\% \\
\cmidrule(l){3-6}
& & End-to-End Latency (s) & \multicolumn{3}{c}{42.0} \\
\bottomrule
\end{tabular}
\end{table*}

Table \ref{tab2} presents the performance of Hyperion with LLaMA 3 and Phi-3 models on three-tier network with 128 output tokens per request. First, the Phi-3-medium model has higher resource utilization than LLaMA 3-8B, exhibiting higher end-to-end latency and showing higher GPU and memory utilization in the reported configurations. For instance, under the 1 Gbps condition, LLaMA 3's latency was 37.5\% lower than Phi-3-medium's (24.8s vs. 39.7s), while Phi-3-medium's memory utilization on the AGX Orin was higher (69.2\% vs. 42.7\%).

As computing capacity decreases from the AGX Orin to the Orin Nano, both average GPU and memory utilization consistently increase for any given model and bandwidth. This is consistent with the smaller capacities of the lower-tier devices. For example, with LLaMA 3 at 1 Gbps, GPU utilization rises from 45.9\% on AGX Orin to 52.1\% on Orin Nano. Reducing the bandwidth from 1 Gbps to 100 Mbps increased end-to-end latency by 9.7\% for LLaMA 3 and 5.8\% for Phi-3-medium.

\subsection{Resource Utilization}

\begin{figure}[t]
    \centering
    \includegraphics[width=0.9\linewidth]{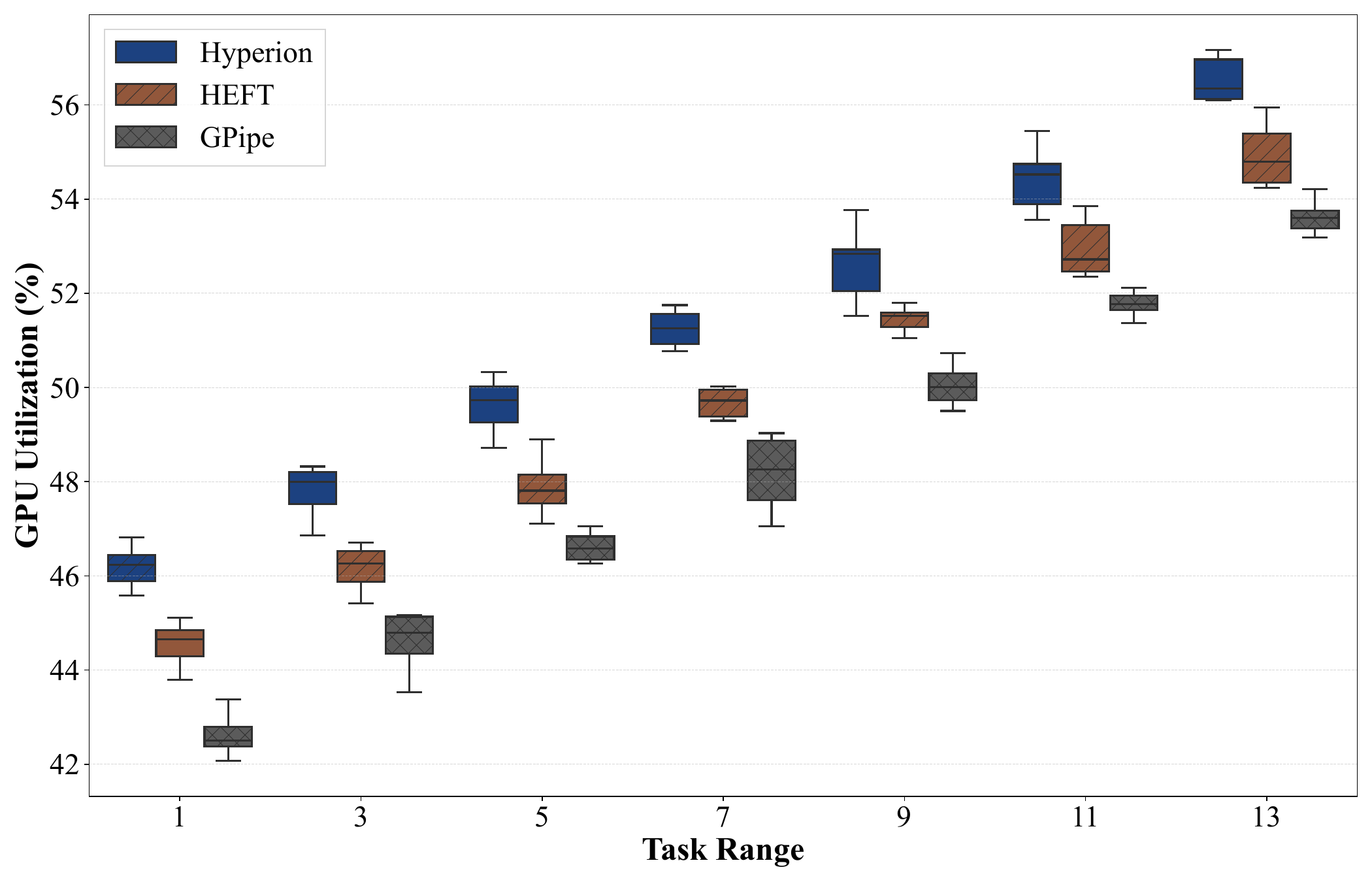} 
    \caption{GPU Utilization on the AGX Orin under Increasing Task Loads.} \label{test3}
\end{figure}

We evaluated the GPU utilization of three algorithms, exemplified on the AGX Orin nodes. With results illustrated in Fig. \ref{test3}, at 3 tasks, Hyperion exhibited the highest median GPU utilization at approximately 47.8\%. HEFT showed a slightly lower median usage of about 46.3\%, while GPipe had the lowest at around 44.9\%. At 13 tasks, GPU utilization increased for all three algorithms. Hyperion’s median utilization rose to approximately 56.3\%. At this point, the median GPU utilization of HEFT and GPipe also increased, to about 55.0\% and 53.4\% respectively, but remained lower than that of Hyperion.

\begin{figure}[t]
    \centering
    \includegraphics[width=1.01\linewidth]{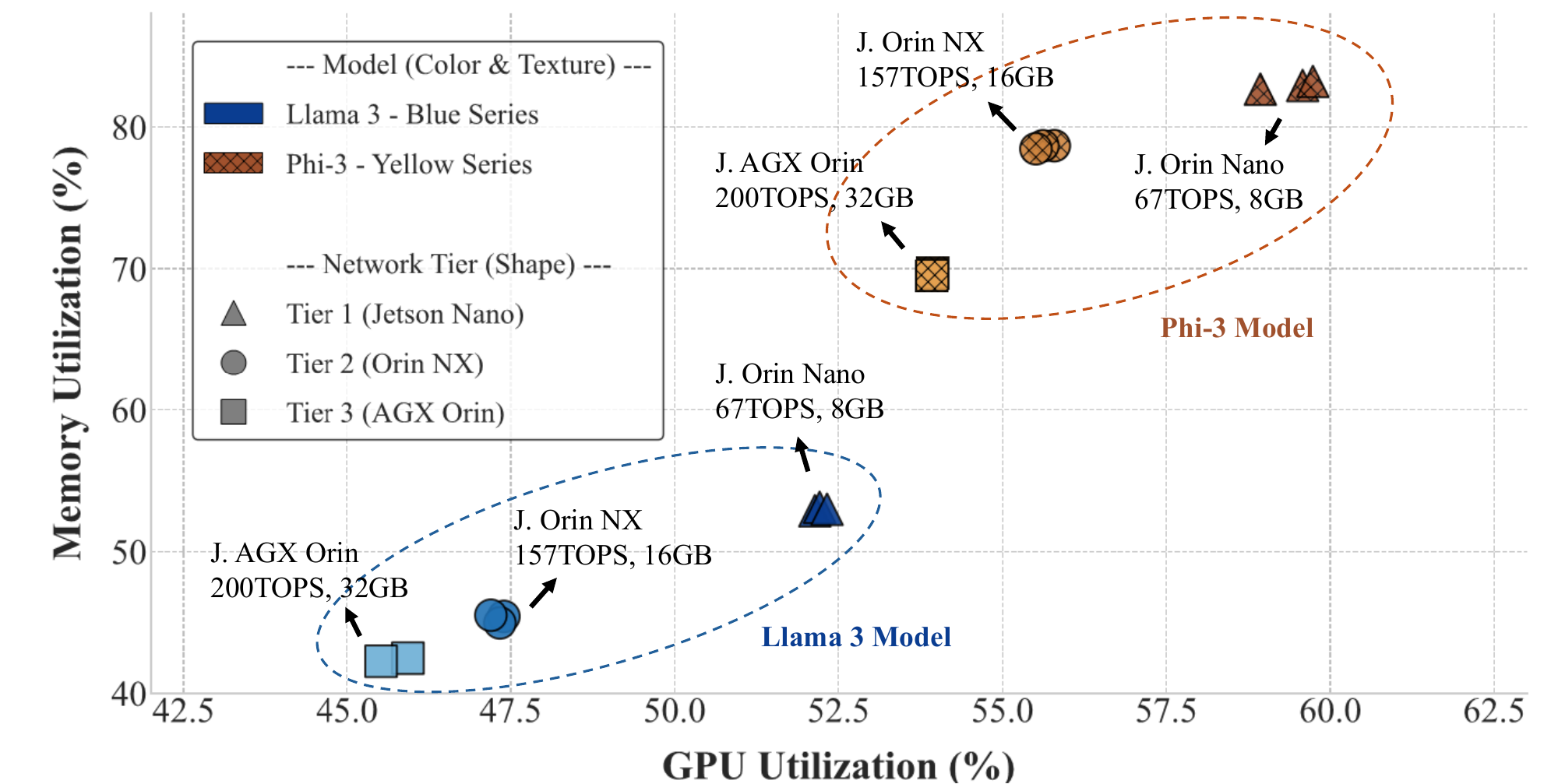} 
    \caption{GPU and memory utilization by tier for Llama 3 and Phi-3-medium under Hyperion.} \label{test4}
\end{figure}

Next, we record the resource utilization of each tier device when running LLaMA 3, with 128 output tokens per request. As shown in Fig. \ref{test4}, devices in Tier 1 demonstrated the most intensive resource usage, with GPU utilization clustered between approximately 51.2\% and 53.5\% and corresponding memory utilization ranging from 51.6\% to 53.8\%. Following this, Tier 2 nodes operated at an intermediate level, exhibiting GPU utilization in the range of 46.1\% to 47.5\% and memory utilization between 45.1\% and 46.5\%. Tier 3 nodes showed the lowest consumption, with GPU utilization around 45.9\% and memory utilization consistently around 42.7\%.

Analysis of the Phi model's tier wise resource utilization also reveals a clear hierarchical structure. GPU utilization was distinctly stratified, with Tier 1 averaging approximately 59.0\%, followed by Tier 2 at 55.3-56.0\% and Tier 3 at 53.5-54.0\%. This hierarchical pattern was even more pronounced in memory consumption, where Tier 1 reached roughly 83.0\%, Tier 2 clustered around 78.0-79.0\%, and Tier 3 used the least at approximately 69.0\%. The values show a consistent tier-wise utilization pattern in this experiment. The higher utilization observed on the lower-capability tiers is consistent with their smaller computing and memory capacities.

\subsection{Latency with Increasing Output Length}

\begin{figure}[t]
    \centering
    \begin{minipage}[b]{0.9\linewidth}
        \centering
        \includegraphics[width=\linewidth]{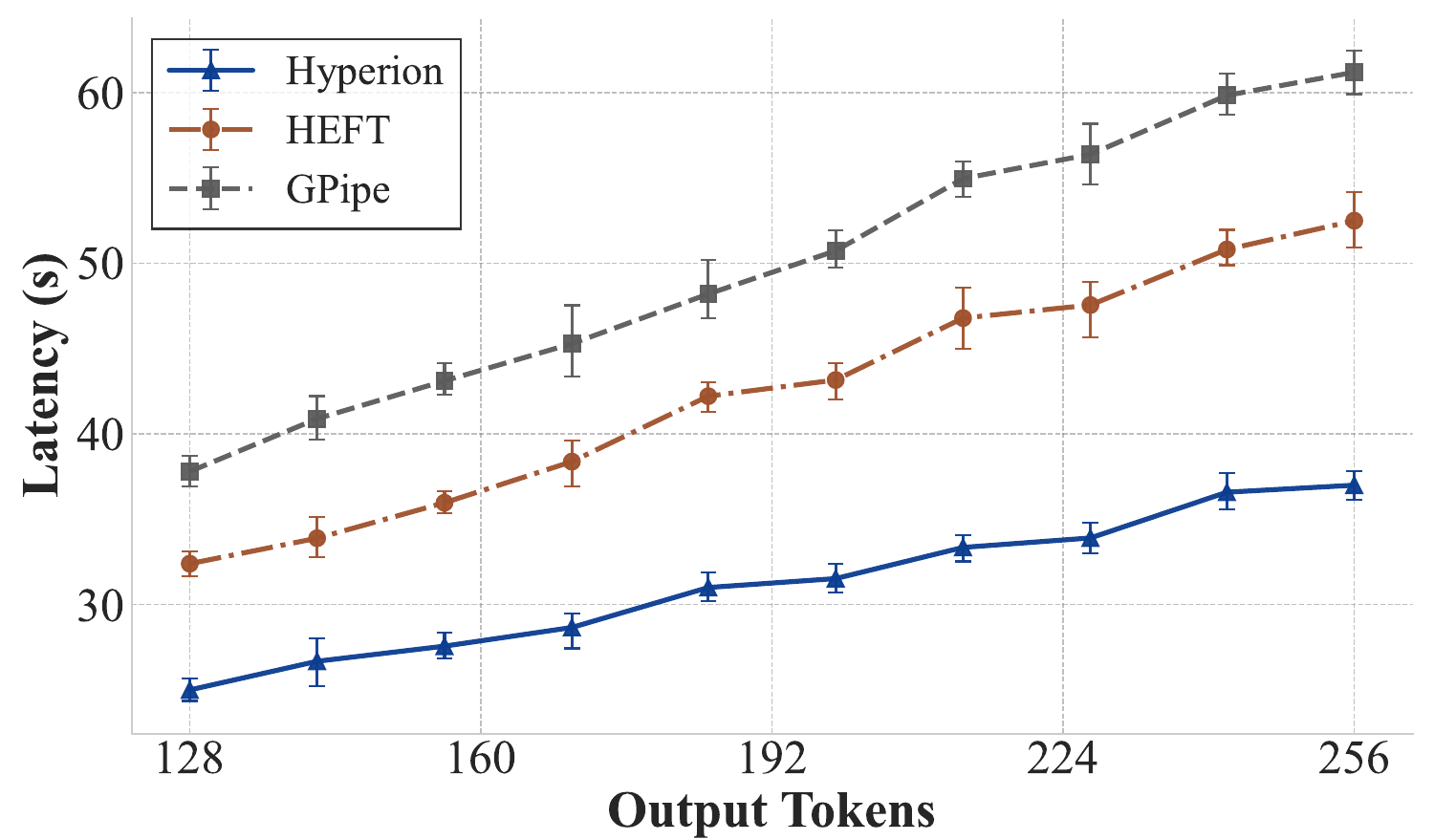}
        \subcaption{Llama 3} 
    \end{minipage}
    \vskip\baselineskip
    \begin{minipage}[b]{0.9\linewidth}
        \centering
        \includegraphics[width=\linewidth]{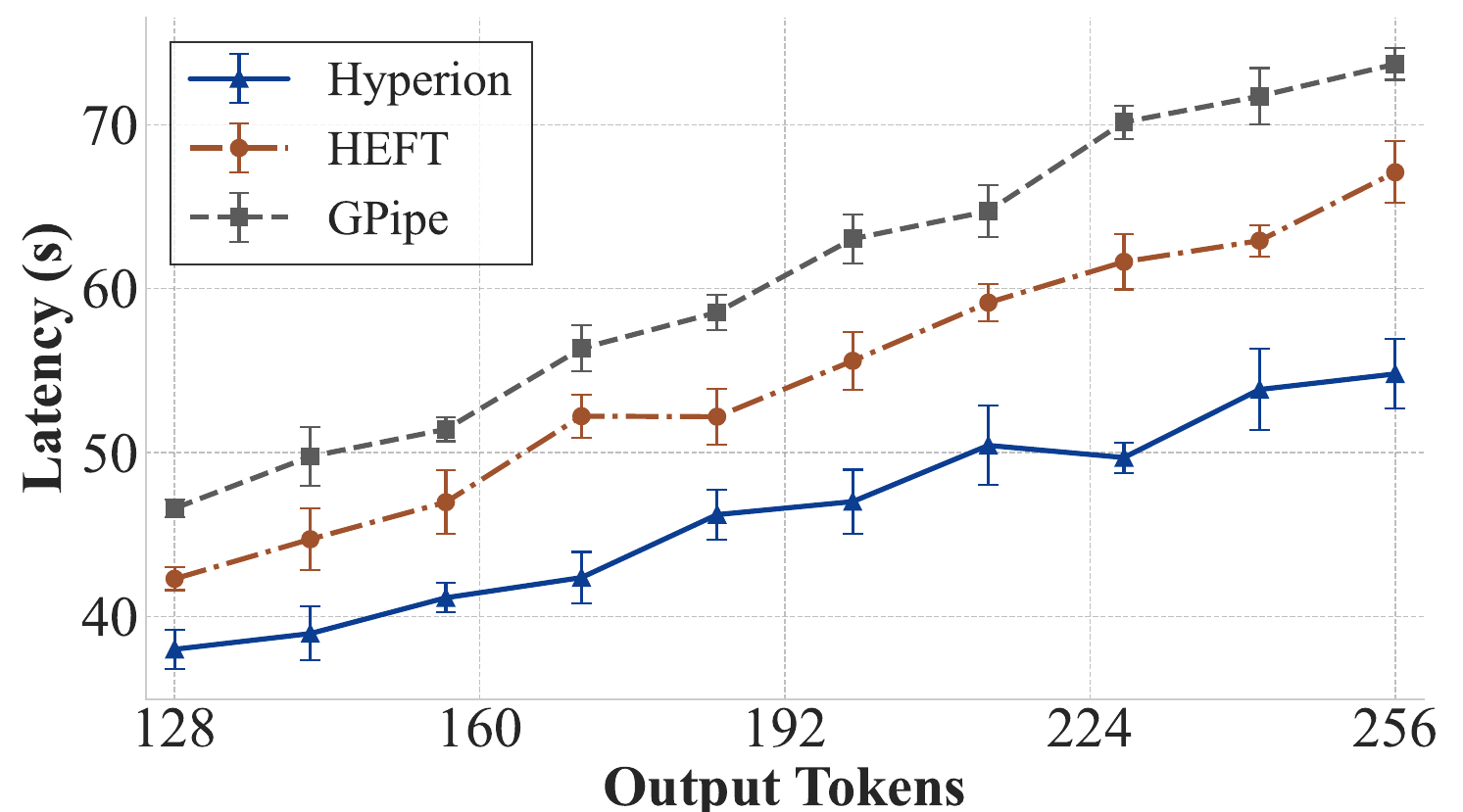}
        \subcaption{Phi-3} 
    \end{minipage}
    \caption{End-to-End Latency Comparison of Hyperion, HEFT, and GPipe across Different Token Output for (a) Llama 3 and (b) Phi-3.} \label{test5}
\end{figure}

Next, we test the performance difference between Hyperion and the baseline algorithm under different output lengths. As shown in Fig. \ref{test5} (a), GPipe consistently demonstrates the highest latency, starting at 37.7s for 128 tokens and increasing to about 61.2s at 256 tokens. The HEFT algorithm performs slightly better but follows a similar increase, with its latency increasing from 32.3s to 52.4s over the same token range. Hyperion records the lowest latency among the three compared methods. It begins with the lowest latency of approximately 24.9s at 128 tokens and reaches 36.9s at 256 tokens.

For Phi-3-medium in Fig. \ref{test5}(b), latency increases with the number of output tokens for all three algorithms. At 128 output tokens, the measured latencies are 38.1s for Hyperion, 42.3s for HEFT, and 46.6s for GPipe. At 256 output tokens, they increase to 54.8s, 67.0s, and 73.6s, respectively. Thus, at 256 output tokens in this setting, Hyperion achieves 18.2\% lower latency than HEFT and 25.5\% lower latency than GPipe. The widening difference suggests that Hyperion scales more favorably with output length than these two baselines under the tested configuration.

\begin{figure}[t]
    \centering
    \includegraphics[width=0.9\linewidth]{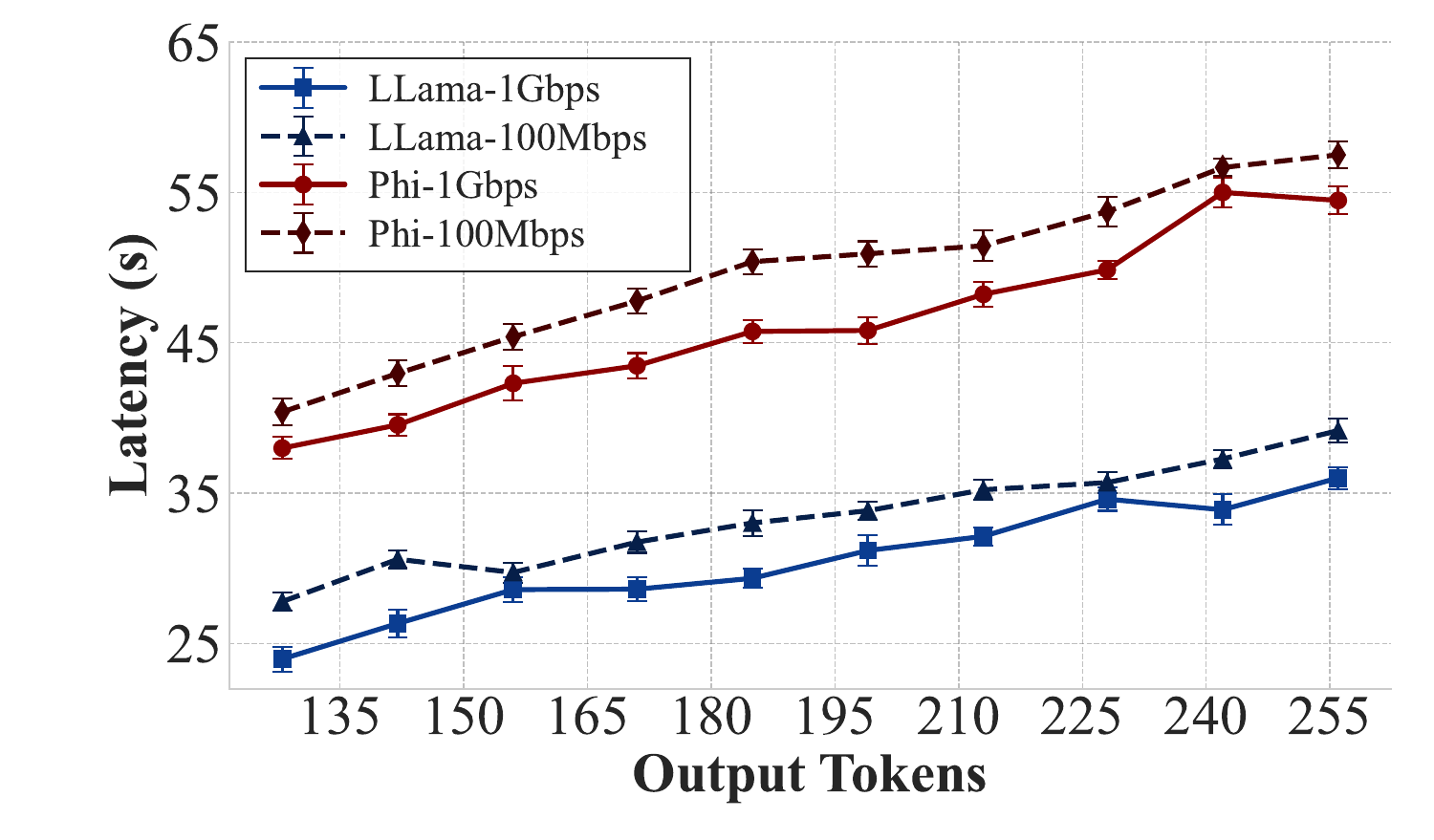} 
    \caption{Impact of Bandwidth Variation on the Performance of the Hyperion across Different Output Token.} \label{test6}
\end{figure}

Fig. \ref{test6} shows the performance of Hyperion under different models and bandwidths. As shown in the figure, there is a positive correlation between latency and the number of tokens. Specifically, under a 1 Gbps bandwidth condition (solid line), its latency for Phi-3 grows from approximately 37.9s at 128 tokens to 54.4s at 256 tokens. The 100 Mbps condition (dashed line) for Phi-3 follows a similar increasing trend, starting slightly higher at 40.3s and reaching 57.4s. In contrast, Llama 3 model, represented by lines with blue, demonstrates substantially lower latency. For the 1Gbps case (solid line), Llama 3's latency increases from 24s to 36s across the same token range. At 100 Mbps (dashed line), the latency follows a similar trend, rising from 28 s to 39 s. Within the evaluated 100 Mbps--1 Gbps range, increasing the effective link rate has a smaller effect on end-to-end latency than increasing the output length. The measured difference between the two link-rate settings is approximately 2-5s over the evaluated token range.

\begin{figure}[t]
    \centering
    \begin{minipage}[b]{0.9\linewidth}
        \centering
        \includegraphics[width=\linewidth]{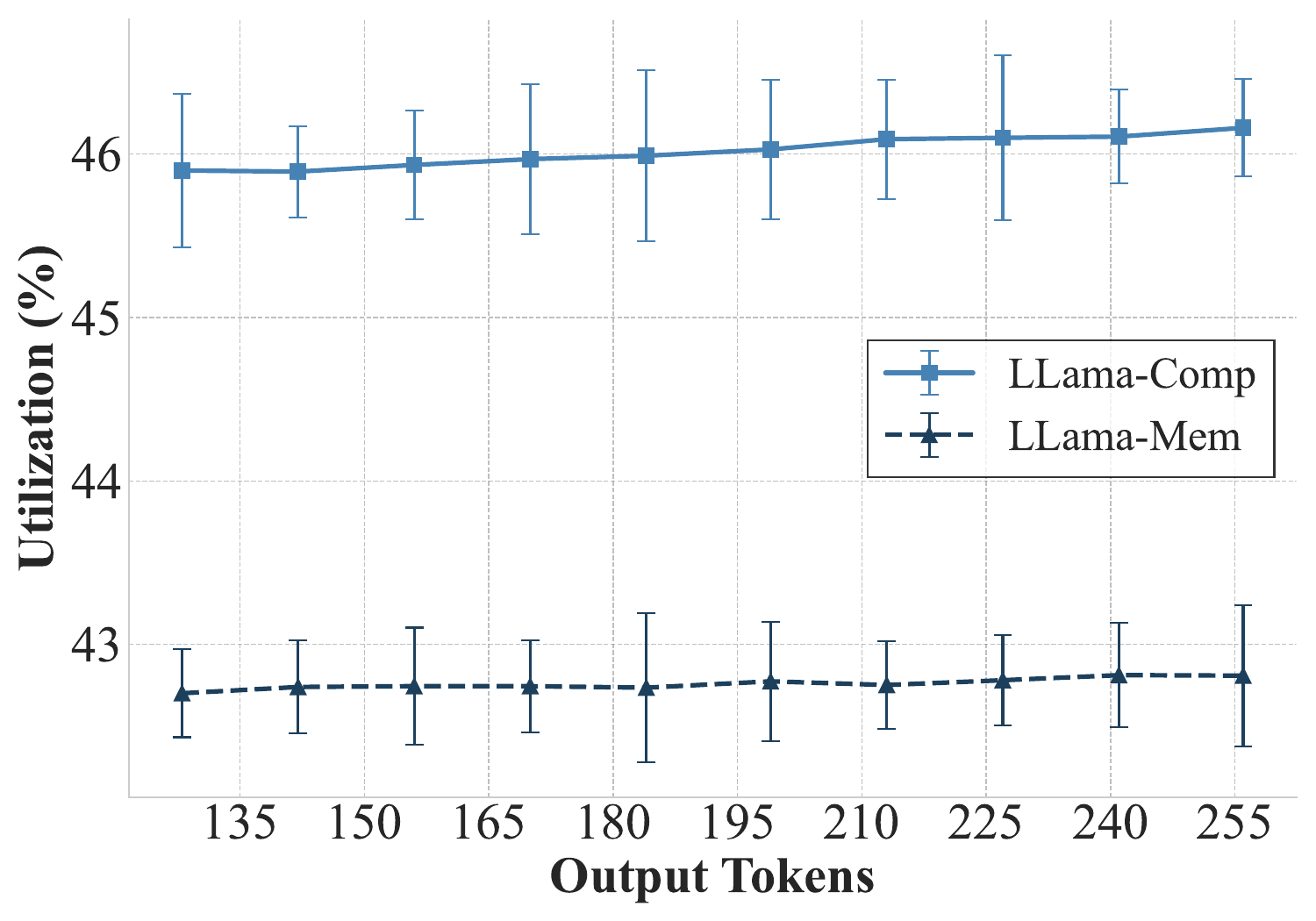}
        \subcaption{Llama 3}
    \end{minipage}
    \vskip\baselineskip
    \begin{minipage}[b]{0.9\linewidth}
        \centering
        \includegraphics[width=\linewidth]{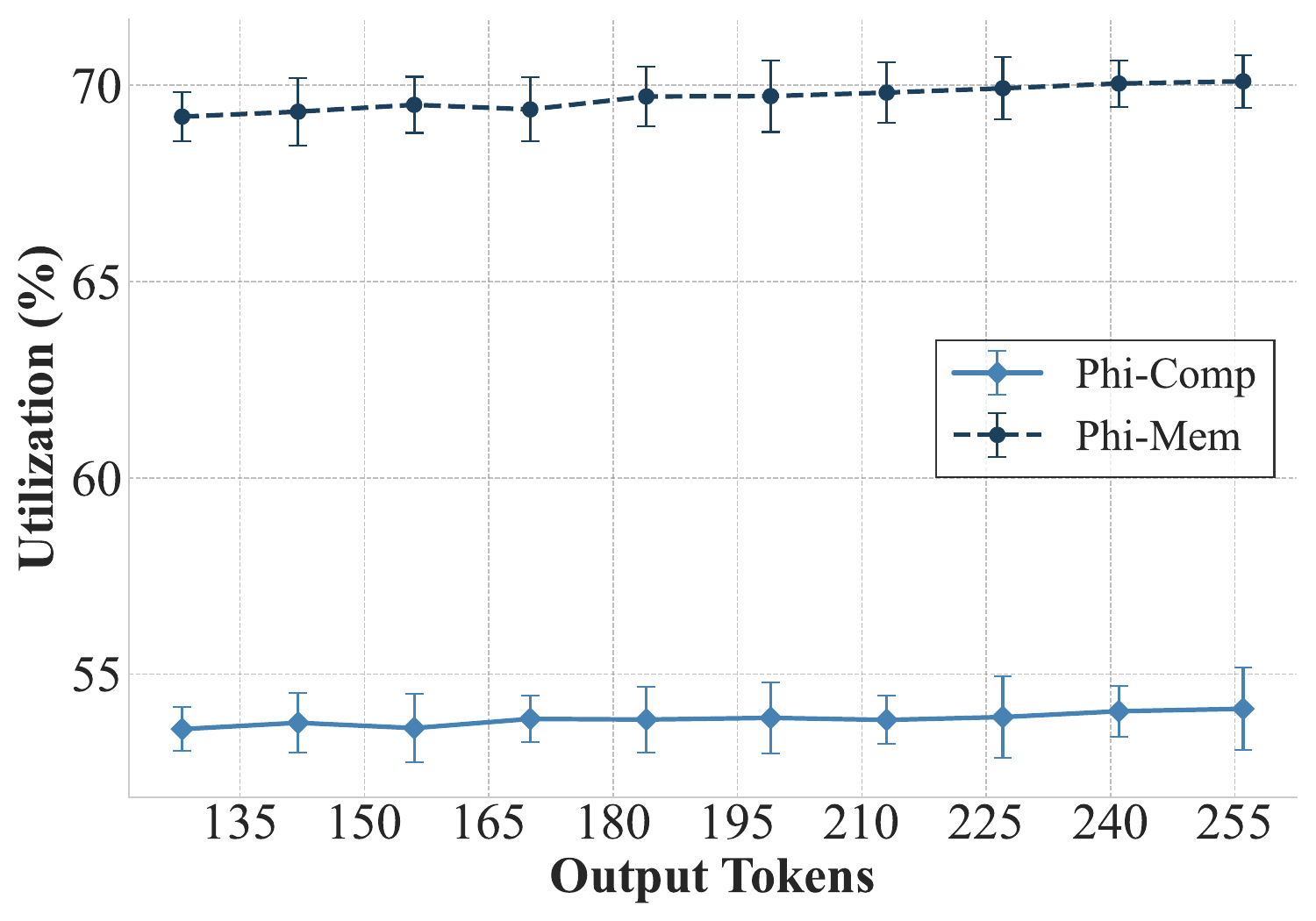}
        \subcaption{Phi-3}
    \end{minipage}
    \caption{Analysis of Llama 3 and Phi-3's Resource Utilization on AGX Orin with Increasing Token.} \label{test7}
\end{figure}

Fig. \ref{test7} (a) reports GPU and memory utilization on the AGX Orin under Hyperion. GPU and memory utilization increase slightly with output length. As shown in Fig. \ref{test7} (a), the computing utilization, represented by a solid line with light blue, exhibits a positive correlation with the token count. It begins at approximately 45.9\% utilization for 128 tokens and rises to 46.2\% at 256 tokens. Similarly, the memory utilization slowly rises from approximately 42.7\% as the number of tokens being generated increases.

Next we use the larger Phi-3 for statistics shown in Fig. \ref{test7} (b). GPU utilization remains between approximately 53.5\% and 54.1\% across the tested output lengths. In contrast, memory utilization increases slightly with output length. Starting from around 69.2\% at 128 tokens, it fluctuates as the token count increases, reaching approximately 70.1\% at 256 tokens.

\subsection{Performance across Different Network Architectures}

\begin{table}[t]
\centering
\caption{Definition of Heterogeneous Network Topologies for Scalability Evaluation.}
\label{tab3}
\begin{tabular}{l l l c}
\hline
\textbf{Network Config.} & \textbf{Tier} & \textbf{Device} & \textbf{Quantity} \\
\hline
\multirow{2}{*}{Two-Tier} & Tier 1 & J. Orin NX & 3 \\
                          & Tier 2 & J. AGX Orin & 2 \\
\hline
\multirow{3}{*}{Three-Tier} & Tier 1 & J. Orin Nano & 3 \\
                           & Tier 2 & J. Orin NX & 3 \\
                           & Tier 3 & J. AGX Orin & 2 \\
\hline
\multirow{4}{*}{Four-Tier} & Tier 1 & J. Orin Nano & 2 \\
                           & Tier 2 & J. Orin Nano & 2 \\
                           & Tier 3 & J. Orin NX & 3 \\
                           & Tier 4 & J. AGX Orin & 3 \\
\hline
\end{tabular}
\end{table}

\begin{figure}[htp]
    \centering
    \begin{minipage}[b]{0.9\linewidth}
        \centering
        \includegraphics[width=\linewidth]{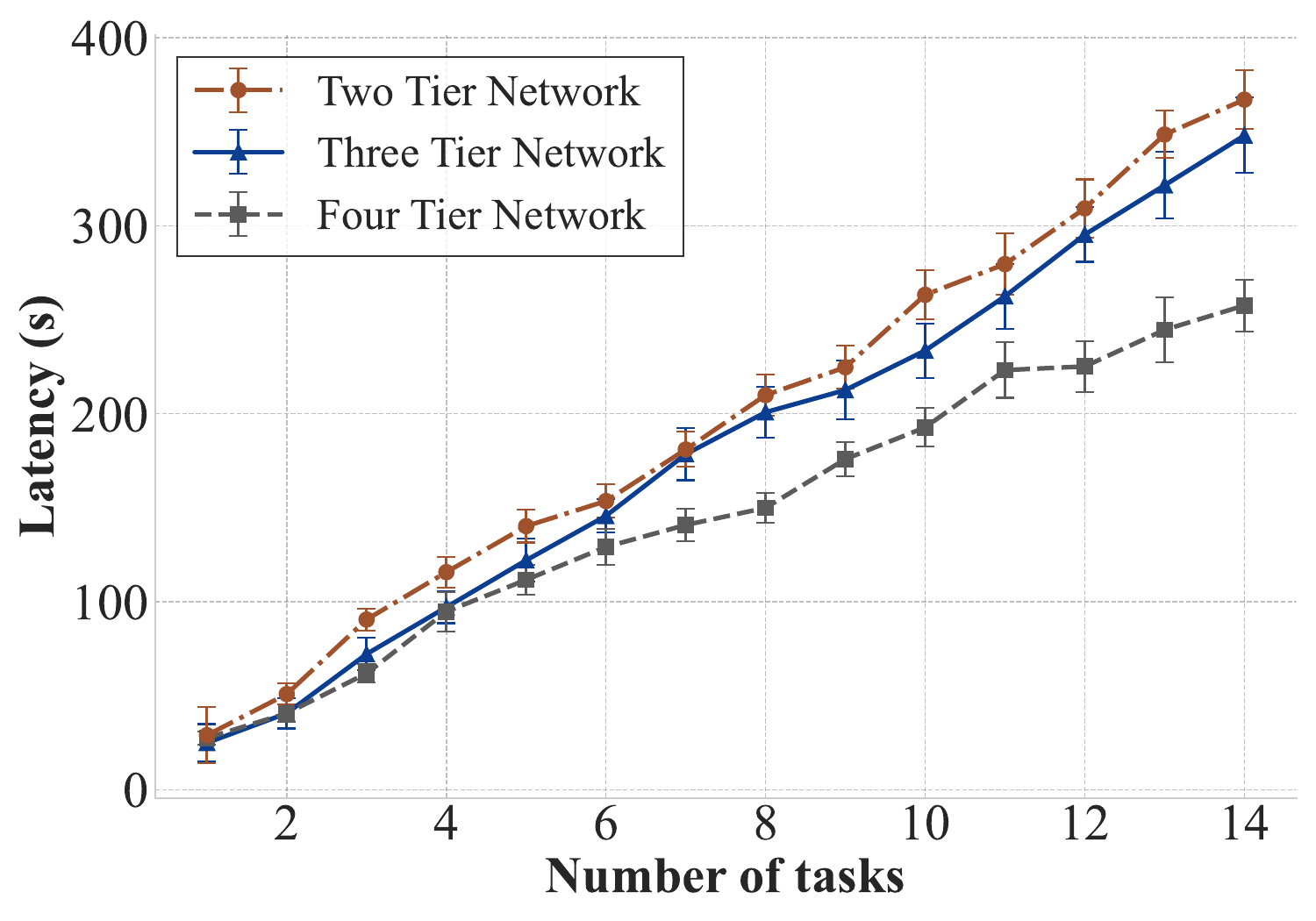}
        \subcaption{Llama 3} 
    \end{minipage}
    \vskip\baselineskip
    \begin{minipage}[b]{0.9\linewidth}
        \centering
        \includegraphics[width=\linewidth]{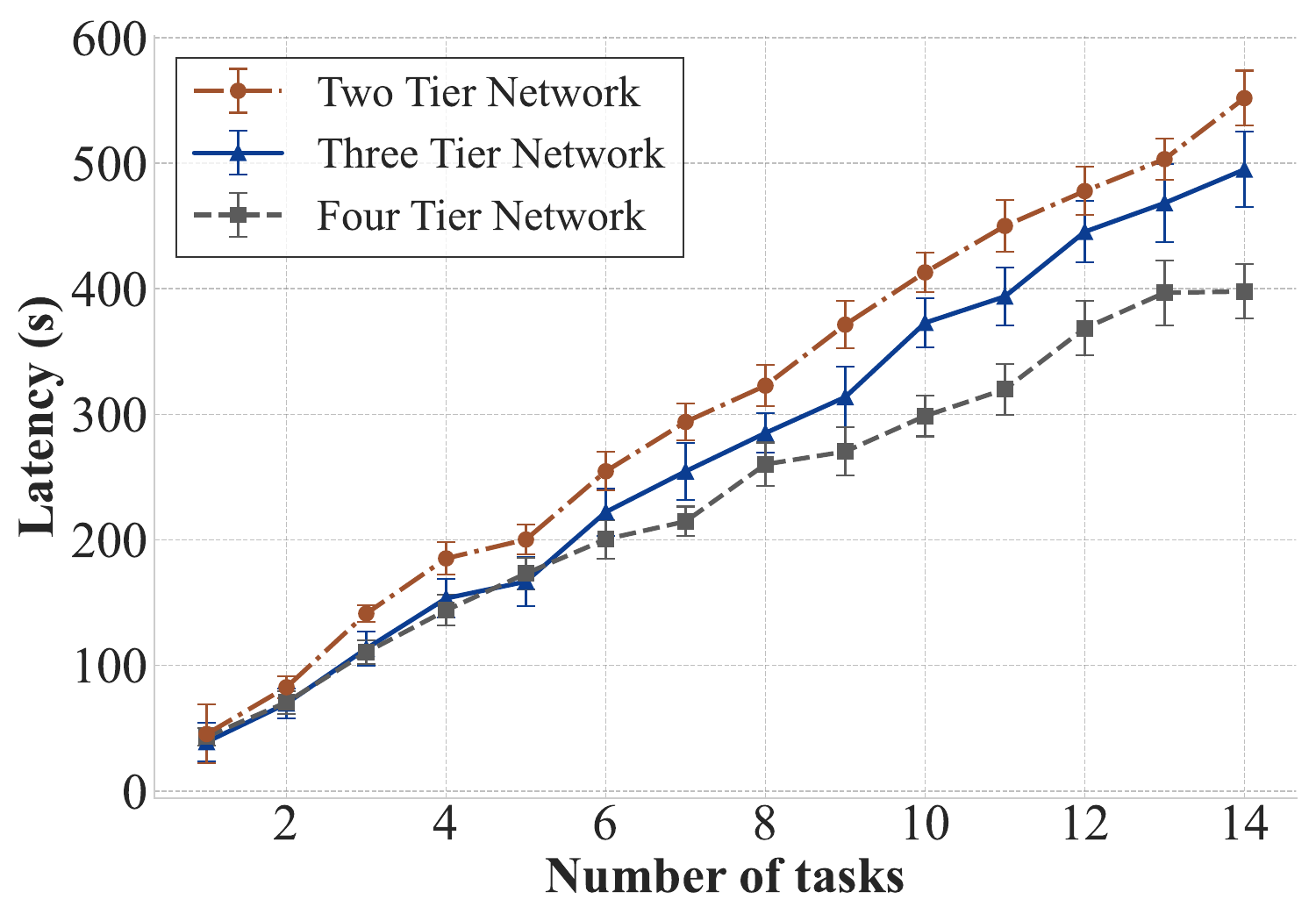}
        \subcaption{Phi-3} 
    \end{minipage}
    \caption{Performance of Hyperion for Llama 3 and Phi-3 Deployment Across Various Network Architectures.} \label{test8}
\end{figure}

Next, we evaluate the performance of Hyperion on different network topologies. The network configuration for each case is shown in Table \ref{tab3}. The network structure becomes progressively more complex, with the number of tiers increasing from two to four, and the total amount of resources gradually expands. Within each configuration, earlier tiers use devices with lower computing capacity.

Fig. \ref{test8} (a) illustrates the latency across different network architectures as a function of the number of tasks. At 6 tasks, the latencies for the four tier, three tier, and two tier networks were approximately 129s, 146s, and 154s, respectively. As the number of tasks increased to 14, the four tier network's latency reached approximately 258s, while the three tier and two tier networks experienced higher latencies of approximately 348s and 367s, respectively. This indicates that the three tier and two tier networks exhibited latencies approximately 35.1\% and 42.4\% higher than the four tier network, respectively. 

For Phi-3-medium in Fig. \ref{test8} (b), the latency difference among the evaluated network configurations becomes larger as the task count increases. At four tasks, the measured latencies of the four-tier, three-tier, and two-tier networks are approximately 143.9s, 153.5s, and 185.3s, respectively. At 14 tasks, the corresponding latencies are approximately 397.8s, 494.9s, and 551.8s. In Fig. \ref{test8}, the four-tier network records the lowest latency for both models at higher task loads.

\section{Conclusion}

This paper studied end-to-end latency for pipelined LLM inference in heterogeneous, memory-constrained multi-tier networks. Hyperion combines offline, resource-aware stage partitioning through HypSplit-DP with online, state-aware node selection through HypSched-RT. The offline stage uses profiled tier capacities to determine memory-feasible contiguous stage boundaries, while the online stage dispatches each request using current node availability and remaining queued workload.

Across the evaluated models, task loads, bandwidth settings, arrival patterns, and output lengths, Hyperion records lower end-to-end latency than the compared baselines. For LLaMA 3 at 14 tasks and 1 Gbps, Hyperion achieves a latency of approximately 355s, which is 31.1\% and 46.9\% lower than the HEFT-based and GPipe-based baselines, respectively. For Phi-3-medium at 256 output tokens, Hyperion achieves a latency of 54.8s, which is 25.5\% lower than the GPipe-based baseline. These results indicate that coordinating offline partitioning with real-time, state-aware node selection can improve latency in the evaluated multi-tier environments.

\bibliographystyle{IEEEtran}
\bibliography{hyperion_references}

\vspace{11pt}

\vspace{11pt}

\vfill

\end{document}